\documentclass[a4paper,oneside,12pt]{article}

\usepackage{epsfig,array,amsmath,amssymb,psfrag,graphicx,tensor,color}
\definecolor{c1}{RGB}{100, 0, 200}
\definecolor{blue6}{RGB}{0, 0, 200} 

\usepackage[colorlinks=true, linkcolor=blue6, citecolor=blue6, urlcolor=c1]{hyperref}

\newcommand{\beq}{\begin{equation}}
\newcommand{\eeq}{\end{equation}}
\newcommand{\bea}{\begin{eqnarray}}
\newcommand{\eea}{\end{eqnarray}}

\newcommand{\CC}{\mathbb{C}}

\newcommand{\NN}{\mathbb{N}}

\newcommand{\mc}[1]{\mathcal{#1}}

\newcommand{\mbb}[1]{\mathbb{#1}}

\newcommand{\ii}{\mathrm{i}}

\makeatletter \@addtoreset{equation}{section} \makeatother 
\renewcommand{\theequation}{\arabic{section}.\arabic{equation}}

\begin{document}

\title{Noether's theorems and conserved currents in gauge theories in the presence of fixed fields}

\author{
G\'abor Zsolt T\'oth
\\[4mm] 
\small \textit{Institute for Particle and Nuclear Physics, Wigner RCP,} \\
\small \textit{MTA Lend\"ulet Holographic QFT Group, Konkoly Thege Mikl\'os \'ut 29-33,} \\
\small \textit{1121 Budapest, Hungary} \\
\small \texttt{toth.gabor.zsolt@wigner.mta.hu} \\
\date{}
}
\maketitle

\begin{abstract}
We extend the standard construction of conserved currents for matter fields in general relativity to general gauge theories. 
In the original construction the conserved current associated with a spacetime symmetry generated by a Killing field $h^\mu$ is given by $\sqrt{-g}\,T^{\mu\nu}h_\nu$, where $T^{\mu\nu}$ is the energy-momentum tensor of the matter.
We show that if in a Lagrangian field theory that has gauge symmetry in the general Noetherian 
sense some of the elementary fields are fixed 
and are invariant under a particular infinitesimal
gauge transformation, then there is a current $\mc{B}^\mu$ that is analogous to 
$\sqrt{-g}\,T^{\mu\nu}h_\nu$
and is conserved if the non-fixed fields satisfy their Euler--Lagrange equations. The conservation of $\mc{B}^\mu$ can be seen as a consequence of an identity that is a generalization of $\nabla_\mu T^{\mu\nu}=0$ and is a consequence of the gauge symmetry of the  Lagrangian. This identity holds in any configuration of the fixed fields if the non-fixed fields satisfy their Euler--Lagrange equations.
We also show that $\mc{B}^\mu$ differs from the relevant canonical Noether current by the sum of an identically conserved current and a term that vanishes if the non-fixed fields are on-shell. 
As example we discuss the case of general, possibly fermionic, matter fields propagating in fixed gravitational and Yang--Mills background. 
We find that in this case the generalization of $\nabla_\mu T^{\mu\nu}=0$ is the Lorentz law 
$\nabla_\mu T^{\mu\nu} - F^{a\nu\lambda}\mc{J}_{a\lambda} = 0$, which holds as a consequence of the 
diffeomorphism, local Lorentz and Yang--Mills gauge symmetry of the matter Lagrangian. 
As a second simple example we consider the case of general fields propagating in a background that consists of a gravitational and a real scalar field.
\end{abstract}

\vspace{3cm}
Journal version: \href{https://doi.org/10.1103/PhysRevD.96.025018}{Physical Review D {\bf 96}, 025018 (2017)}

\maketitle

\thispagestyle{empty}

\newpage



\section{Introduction}
\label{sec.intr}

In general relativity, the most usual way to construct conserved currents associated with spacetime symmetries
for matter fields is to contract the
Einstein--Hilbert energy-momentum tensor $T^{\mu\nu}$ with the Killing vector field $h^\mu$ that generates the relevant symmetry. It is easy to verify that the current $j^\mu = \sqrt{-g}\, T^{\mu\nu}h_\nu$ (where $g$ denotes the determinant of the metric)
obtained in this way is conserved (i.e.\ $\partial_\mu j^\mu=0$) 
as a consequence of the Killing equation $\nabla_\mu h_\nu + \nabla_\nu h_\mu=0$ 
and of the divergencelessness of $T^{\mu\nu}$ ($\nabla_\mu T^{\mu\nu} = 0$). The latter property of $T^{\mu\nu}$ 
is ensured if the matter Lagrangian density transforms as a scalar density under diffeomorphisms and the matter fields satisfy their equations of motion (see e.g.\ Section E.1 of \cite{WaldGR}). 
The metric does not need to satisfy its field equations, i.e.\ it can be a fixed, external field. 
On the other hand, Noether's first theorem can also be used to construct a conserved current associated with $h^\mu$, if the matter fields admit a Lagrangian description.
This current generally differs from $\sqrt{-g}\, T^{\mu\nu}h_\nu$ and its construction is also apparently completely different. Nevertheless, it can be shown that
if the matter fields satisfy their Euler--Lagrange (EL) equations,
then the difference between the two currents is an identically conserved current, i.e.\ a current of the form $\partial_\mu \Sigma^{\mu\nu}$, 
where $\Sigma^{\mu\nu}$ is antisymmetric \cite{IW, Szabados}.

To some extent similarly, in electrodynamics the electric current is usually defined as the Euler--Lagrange derivative 
of the Lagrangian function with respect to the vector potential, i.e.\ as the source of the electromagnetic field,
but it can also be obtained as the Noether current corresponding to global $U(1)$ gauge transformations.

The primary aim of the present paper is to generalize the construction described above to any gauge theory 
that can be formulated in the framework applied in Noether's classic paper \cite{Noether,Y} on symmetries and conservation laws in Lagrangian field theory. 
This is a very general framework that allows one to study various special kinds of gauge theories, 
for example diffeomorphism covariant theories and Yang--Mills (YM) type gauge theories, on the same footing.   

Our main motivation, besides general interest in conservation laws in gauge theories, to look for a generalization 
of the above construction is that recent developments \cite{NQV,DM1,DM2,Borokhov} appear to indicate that such a generalization is possible and would be useful for better understanding these developments, for answering certain open questions and for possible further applications.

In \cite{NQV} it was proposed that in the case when some matter propagates in fixed electrically charged black hole background, 
its energy and angular momentum can be obtained using the conserved currents
\beq
\label{eq.0}
j^\mu = \sqrt{-g}\, (T\indices{^\mu_\nu} h^\nu +h^\nu A_\nu \mc{J}^\mu)\ ,
\eeq
where $A_\mu$ is the vector potential of the background electromagnetic field, $\mc{J}^\mu$ is the electric current, 
$T^{\mu\nu}$ is the Einstein--Hilbert energy-momentum tensor of the matter, and $h^\mu$ is the Killing field generating time translations\footnote{See \cite{NQV} regarding AdS black holes.} or rotations. 
It is also assumed here that not only the background electromagnetic field, but also $A_\mu$ is invariant under 
time translations or rotations.  
(\ref{eq.0}) was very useful in \cite{NQV} because, together with a suitable energy condition on $T^{\mu\nu}$, 
it allowed the authors to derive general results for charged black holes without 
more detailed information on the nature of the matter. 
To justify (\ref{eq.0}), it was shown in \cite{NQV} that the conservation of (\ref{eq.0}) can be derived from the 
generalized Lorentz law 
\beq
\label{eq.1}
\nabla_\mu T^{\mu\nu} = F^{\nu\lambda}\mc{J}_\lambda\ ,
\eeq
the Killing equation, 
the symmetry of $T^{\mu\nu}$, the invariance of $A_\mu$, and the conservation of the electric current.
In (\ref{eq.1}) $F^{\mu\nu}$ denotes the electromagnetic field.
A justification of the validity of (\ref{eq.1}) was also given in \cite{NQV}, although it is 
not analogous to the derivation of $\nabla_\mu T^{\mu\nu} = 0$ mentioned in the first paragraph, 
and it uses the Maxwell equation
$\nabla_\mu F^{\mu\nu}=0$ for the background electromagnetic field. 
Nevertheless, a derivation of (\ref{eq.1}) that is analogous to the derivation of $\nabla_\mu T^{\mu\nu} = 0$
is possible, moreover it can be extended to the case of fixed gravitational + YM background \cite{BBGS}. 

In \cite{TG} we also found the current (\ref{eq.0}) for the complex Klein--Gordon field by applying Noether's first theorem, and in \cite{TG2} we found that in the case of the Dirac field as matter field the current (\ref{eq.0}) can be obtained
by adding a certain identically conserved current to the Noether current, if the Dirac equation is satisfied. 
The only information on the background metric (or tetrad) and vector potential that was needed for these results was their invariance under time translation or rotation.

An identity and a current similar to (\ref{eq.1}) and (\ref{eq.0}) also appear in \cite{DM1,DM2}.
In these works (\ref{eq.1}) takes a more general form, 
since some of the currents associated with the fixed covector fields are not conserved. Further comments concerning the formulas appearing in \cite{DM1,DM2} will be made at the end of Section \ref{sec.app1}.
A construction of currents in a general setting that is partly similar to the one in the present paper 
was given in Section 2 of \cite{Borokhov}. This will be discussed further at the end of Section \ref{sec.n4}.  

We note that there is a well-known derivation of the 
generalized Lorentz law in the case when electromagnetic field and matter propagates,
interacting with one another, in a fixed gravitational background (see Section 22.4 of \cite{MTW}).
In this derivation one obtains the generalized Lorentz law from the Maxwell equations and from the divergencelessness of 
the matter + electromagnetic total energy-momentum tensor. The derivation can also be extended to nonabelian gauge theory in a straightforward way. Nevertheless, in this paper we are interested in the more general situation when the electromagnetic (or YM) field is also part of the background. 

Regarding the results mentioned above, several questions can also be raised.
Since in \cite{BBGS} fermionic matter fields were not considered, one would like to derive (\ref{eq.1})
also for the case when some matter fields are fermionic. (\ref{eq.0}) should also be extended to the case of 
gravitational + YM background. One can ask what form (\ref{eq.0}) will take
if the vector potential is invariant 
under the diffeomorphisms generated by $h^\mu$ only up to YM gauge transformation, 
since the latter is a more natural symmetry requirement on the vector potential.   
The difference between (\ref{eq.0}) and the Noether current that one obtains by Noether's first theorem should be investigated as well.

The generalization of the construction described in the first paragraph that we present
is closely related to Noether's theorems and is relatively easy to find
once one has thoroughly understood these theorems, therefore
we review them briefly in a way that is suitable for the purpose of this paper. This review is intended to be quite general but mathematically elementary. 
For expositions of Noether's theorems in the literature and for related results see e.g.\  
\cite{Olver}-\cite{Koivisto}, \cite{IW}, and references therein. Many references can be found in \cite{Y} as well.
 
As an application and illustration of the general construction we discuss first
the example of matter fields coupled to external gravitational 
and YM gauge fields (allowing the electromagnetic field as a special case).
The type of the matter fields and the precise form of their Lagrangian are left unspecified. 
For deriving the generalized Lorentz law and the related currents 
only the symmetry properties of the gravitational and YM gauge fields and of the matter Lagrangian are needed, 
but in other parts for simplicity we make some assumptions on the nature of the matter fields and on the form of the matter Lagrangian. We allow fermionic matter fields, therefore we use tetrad fields as elementary gravitational field variables.
We discuss the cases of the Dirac field and the scalar field as matter field more explicitly,
since these are interesting special cases. 
The aim of this example is also to answer the questions mentioned above.
The second example, which is included on account of its simplicity, 
is the case of arbitrary fields propagating in the presence of external scalar and gravitational fields. 

The paper is organized as follows. Section \ref{sec.n} contains the review of Noether's theorems 
and the generalization of the construction described in the first paragraph. The latter can be found in Section \ref{sec.n4},
which is the central part of the paper.
In Section \ref{sec.eymm} the details of the first example introduced above are given. 
In particular, the generalized Lorentz law 
and the currents (\ref{eq.0}) are discussed in Section \ref{sec.app1}.
The second example is described in Section \ref{sec.5}.
A summary is given in Section \ref{sec.concl}.
Appendix \ref{sec.A} contains definitions and comments related to the examples.

The signature of metric tensors is $(+,-,-,-)$.
The brackets $(\, )$ and $[\, ]$ applied to indices are used to denote symmetrization and antisymmetrization,
and these operations are understood to include division by the number of permutations.
Coordinate based formalism is employed and the notions of  
fiber bundles and differential forms are mostly avoided in order to keep the exposition as elementary as possible. 
The terms gauge theory and gauge symmetry are used in the general sense described in Section \ref{sec.n2},
and it is stated explicitly when a special kind of gauge symmetry is meant.
The coordinate independent YM gauge transformations are called global.

\section{Noether's theorems and the construction of conserved currents in the presence of fixed fields}
\label{sec.n}

In order to describe the situation when some of the 
fields are fixed, i.e.\ do not necessarily satisfy their field equations, we divide
the complete set of elementary fields into two sets. The fields in these sets are denoted as $\Phi_i$ and $\chi_j$, 
where $i$ and $j$ are general indices
labeling the fields and their components. Any of the two sets may be empty.
Generally $\chi_j$ will be the fixed fields, but $\Phi_i$, which may be called dynamical fields, are also not assumed to satisfy their field equations unless explicitly stated. 
The physical role of $\Phi_i$ and $\chi_j$, i.e.\ whether they are matter or other type of fields, is not restricted. 
In particular, the presence of a metric tensor is not required. 
Commuting and anticommuting fields are both allowed.
For simplicity the fields are assumed to be real (self-conjugate) in this section. 
This does not cause any loss of generality, since any complex field is equivalent to two real fields. 
For derivatives with respect to anticommuting variables the following sign convention is used: 
if $\theta$ is an anticommuting variable and $E$ is an expression of the form 
$E_1\theta E_2$, then $\frac{\partial E}{\partial \theta}= (-1)^n E_1 E_2$, where $n=0$ if $E_2$ is even and $n=1$ if 
$E_2$ is odd.

The Lagrangian density function  
$L(x^\mu,  \chi_j, \partial_\mu\chi_j,\partial_{\mu\nu}\chi_j,\dots, \Phi_i, \partial_\mu\Phi_i,\partial_{\mu\nu}\Phi_i,\dots)$
is assumed to be an even 
local function of $\Phi_i$ and $\chi_j$, but otherwise it is allowed to depend on arbitrarily high derivatives. Further assumptions on the Lagrangian are not made in this section and it is not specified what kind of physical system it describes. 
A local function of some fields $\phi_i(x^\mu)$ is defined in this paper to be 
a function of the form $f(x^\mu,  \phi_i(x^\mu), \partial_\nu\phi_i(x^\mu), \partial_{\nu\lambda}\phi_i(x^\mu), \dots)$, 
that depends on the fields and on finitely many derivatives of them 
and may depend explicitly on the coordinates $x^\mu$ as well\footnote{In \cite{Olver} the term 'differential function' is used instead of 'local function'.}.
Although the action integral will not be used, 
it should be noted that the contribution to it from the domain $U$ on which $x^\mu$ are defined is 
$\int_U d^D x\, L$, where $D$ denotes the dimension of the base manifold $M$ on which the fields are defined. The integration measure used here is the measure determined by the coordinate chart. 
The behaviour of $\Phi_i$ and $\chi_j$ under coordinate changes
is not necessary to specify for the purpose of this section. 

It is known that higher derivative theories generally have various undesirable features, particularly instabilities and 
unphysical degrees of freedom, but
they have better renormalizability properties than low derivative theories
and the difficulties caused by the mentioned features may be surmountable
(see e.g.\ \cite{Simon, Woodard, MS, HH, Stelle, MS2, Modesto, Giacchini, Woolliams}).
As far as symmetries are concerned, there is no major reason to restrict the order of the derivatives that may appear in the Lagrangian.

We distinguish three theorems of Noether, and 
within the third theorem we distinguish two parts. 
The first theorem, which is the most well-known, is not specific to gauge theories. 
In the literature the third theorem is often included in the second one, 
but it seems useful to separate them. 
In Noether's paper \cite{Noether,Y} the statement of the 
second theorem does not include the third theorem, but the third theorem does not appear as a 
separate theorem either. 

We do not follow rigorously the original formulation of Noether's theorems. 
In particular, we are interested only in the consequences of symmetries
and do not consider reverse statements. For the latter we refer the reader to the literature, e.g.\ \cite{Olver}.
   
The content of the three theorems can be summarized very briefly as follows.
The first theorem states that if the Lagrangian has a symmetry,
then there exists a current $J^\mu$ (the Noether current) that is conserved (i.e.\ $\partial_\mu J^\mu=0$)
if the fields satisfy their EL equations. 
The second theorem states that if the Lagrangian has a gauge symmetry, 
then the EL derivatives of the Lagrangian with respect to the fields satisfy a differential identity,
thus the EL equations are not independent.
The third theorem states that if the Lagrangian has a gauge symmetry, then the Noether currents associated with these symmetries 
coincide with certain identically conserved currents up to some terms that vanish if all fields satisfy their EL equations.

In Section (\ref{sec.n1}) we present the first theorem in a form that is adapted to the situation when fixed fields are present.
The distinction between $\Phi_i$ and $\chi_j$ is irrelevant in the second and third theorems, 
but we keep it for later use in the construction in Section \ref{sec.n4}. 
The definition of gauge symmetry in general sense can be found in Section \ref{sec.n2}.

After Noether's three theorems we present
the generalization of the construction described in the first paragraph of Section \ref{sec.intr}. 
This generalized construction can also be regarded as an extension of Noether's theorems.
The construction described in Section \ref{sec.intr} has three parts: 
the first one is the construction of the energy-momentum tensor and the derivation of its divergencelessness, 
the second one is the construction of the current from the energy-momentum tensor and from the Killing field
and the derivation of its conservation, 
and the third one is the result that the current constructed in this way 
differs from the Noether current associated with the Killing field in an identically conserved current
if the matter fields satisfy their EL equations. 
Each part is generalized in Section \ref{sec.n4}. 

In general relativity the divergencelessness of the energy-momentum tensor can also be seen as a consequence of the Einstein equation if the metric is not fixed. Similarly, in electrodynamics the conservation of the electric current can be seen as a consequence of the Maxwell equation. 
These observations are also generalized in Section \ref{sec.n4}.  

In the next subsection an auxiliary formula is described, which is of central importance in the subsequent derivations.

\subsection{The partial integration formula}
\label{sec.np}

Let $G$ be a quantity that can be written as
\beq
\label{eq.g1}
G= G_\alpha \epsilon^\alpha + G_\alpha^\nu \partial_\nu \epsilon^\alpha  +  G_\alpha^{\nu\lambda} \partial_{\nu\lambda} \epsilon^\alpha + \dots
\eeq
where $\epsilon^\alpha$ is a function with several components indexed by $\alpha$, 
the functions $G_\alpha^{\nu\lambda}$,  $G_\alpha^{\nu\lambda\rho}$, ... 
are completely symmetric in the upper indices, and the sum on the right hand side contains only finitely many terms.
By straightforward application of the basic differentiation rule $(uv)'=u'v+uv'$ one can show that  
\beq
\label{eq.g2}
G =  \hat{G}_\alpha \epsilon^\alpha + \partial_\nu \mc{G}^\nu\ ,
\eeq
where 
\beq
\label{eq.g3}
\hat{G}_\alpha =G_\alpha -\partial_\nu G^\nu_\alpha +\partial_{\nu\lambda} G^{\nu\lambda}_\alpha - \dots
\eeq
and
\beq
\label{eq.g4}
\mc{G}^\nu= G_\alpha^{\nu}  \epsilon^\alpha  +
(G_\alpha^{\nu\lambda} \partial_\lambda \epsilon^\alpha  - \partial_\lambda  G_\alpha^{\nu\lambda} \epsilon^\alpha)
+( G_\alpha^{\nu\lambda\rho} \partial_{\lambda\rho} \epsilon^\alpha - 
\partial_\rho G_\alpha^{\nu\lambda\rho} \partial_{\lambda} \epsilon^\alpha 
+ \partial_{\lambda\rho} G_\alpha^{\nu\lambda\rho} \epsilon^\alpha)+\dots
\eeq
The $(n+1)$-th term of the sum on the right hand side of (\ref{eq.g3}) is 
$(-1)^n\partial_{\nu_1\nu_2\dots\nu_n}G_\alpha^{\nu_1\nu_2\dots\nu_n}$. The $n$-th group of terms  
on the right hand side of (\ref{eq.g4})  is
\bea
&&  G_\alpha^{\nu\lambda_1\lambda_2\dots\lambda_{n-1}} \partial_{\lambda_1\lambda_2\dots\lambda_{n-1}}\epsilon^\alpha
- \partial_{\lambda_{n-1}} G_\alpha^{\nu\lambda_1\lambda_2\dots\lambda_{n-1}} \partial_{\lambda_1\lambda_2\dots\lambda_{n-2}}\epsilon^\alpha\nonumber\\
&&\qquad + \partial_{\lambda_{n-2}\lambda_{n-1}}G_\alpha^{\nu\lambda_1\lambda_2\dots\lambda_{n-1}} \partial_{\lambda_1\lambda_2\dots\lambda_{n-3}}\epsilon^\alpha - \dots\nonumber\\
&&\qquad\qquad + (-1)^{n-1} \partial_{\lambda_1\lambda_2\dots\lambda_{n-1}}G_\alpha^{\nu\lambda_1\lambda_2\dots\lambda_{n-1}} \epsilon^\alpha\ .
\nonumber\eea
Since the sum on the right hand side of (\ref{eq.g1}) is finite, the sums in (\ref{eq.g3}) and  (\ref{eq.g4}) are also finite.
Moreover, if 
$G_\alpha$, $G_\alpha^\nu$, $G_\alpha^{\nu\lambda}$, ... do not depend on $\epsilon^\alpha$ (and on its derivatives), 
then if all terms beyond the first $n$ terms are zero on the right hand side of (\ref{eq.g1}), 
then $\mc{G}^\nu$ does not depend on higher than $(n-2)$-th derivatives of $\epsilon^\alpha$.

We call (\ref{eq.g2}) the partial integration formula, since it can be regarded as a generalization of the 
differentiation rule $(uv)'=u'v+uv'$ on which partial integration is based. 
It is clear that adding an arbitrary conserved current to $\mc{G}^\nu$ preserves
(\ref{eq.g2}), but we always use the definition (\ref{eq.g4}).

\subsection{The first theorem}
\label{sec.n1}

A one-parameter transformation of the fields can be written after linearization in the parameter, denoted by $s$, as
\beq
\label{eq.tr1}
\Phi_i\to \Phi_i+s\delta \Phi_i\ ,\qquad 
\chi_j\to \chi_j+s\delta \chi_j\ .
\eeq
$s$ is assumed to be real number valued and $\delta \Phi_i$ and $\delta \chi_j$ are assumed to have the same commutation character 
as $\Phi_i$ and $\chi_j$, respectively.
Generally both $\delta \Phi_i$ and $\delta \chi_j$ may depend on  
$\Phi_i$ and $\chi_j$ and on their derivatives. 
Such a transformation may be induced by a transformation in the base manifold or in the target space of the fields, but may be more general. Supersymmetry transformations, for example, also fit in this framework.
The associated first order variation of $L$ is defined as
$ \delta L = \frac{dL}{ds}|_{s=0}  = \delta L_\Phi + \delta L_\chi$ with
\bea
\delta L_\Phi & = & \frac{\partial L}{\partial \Phi_i}\delta\Phi_i +
\frac{\partial L}{\partial (\partial_\mu\Phi_i)}\partial_\mu\delta \Phi_i +
\frac{\partial L}{\partial (\partial_{\mu\nu}\Phi_i)}\partial_{\mu\nu} \delta \Phi_i + \dots 
\label{eq.dl1} \\
\delta L_\chi & = & \frac{\partial L}{\partial \chi_j}\delta\chi_j +
\frac{\partial L}{\partial (\partial_\mu\chi_j)}\partial_\mu\delta \chi_j +
\frac{\partial L}{\partial (\partial_{\mu\nu}\chi_j)}\partial_{\mu\nu} \delta \chi_j + \dots 
\label{eq.dl2}
\eea
Applying the partial integration formula (\ref{eq.g2}) to (\ref{eq.dl1}) with
$\alpha\to i$, $\epsilon^\alpha\to \delta\Phi_i$ and
$G^{\mu\nu\dots}_\alpha\to\frac{\partial L}{\partial (\partial_{\mu\nu\dots}\Phi_i)}$
gives
\beq
\label{eq.dl3}
\delta L_\Phi  =  \frac{\delta L}{\delta \Phi_i}\delta\Phi_i 
+ \partial_\mu j_\Phi^\mu\ , 
\eeq
where 
\beq
\label{eq.el1}
\frac{\delta L}{\delta \Phi_i} =
\frac{\partial L}{\partial \Phi_i}-\partial_\mu\frac{\partial L}{\partial (\partial_\mu\Phi_i)}
+\partial_{\mu\nu}\frac{\partial L}{\partial (\partial_{\mu\nu}\Phi_i)}-
\partial_{\mu\nu\lambda}\frac{\partial L}{\partial (\partial_{\mu\nu\lambda}\Phi_i)}+
\dots
\eeq
and
\bea
j_\Phi^\mu & = &  \frac{\partial L}{\partial (\partial_\mu\Phi_i)}\delta \Phi_i+
\biggl(\frac{\partial L}{\partial (\partial_{\mu\nu}\Phi_i)}\partial_\nu \delta \Phi_i
-\partial_\nu\frac{\partial L}{\partial (\partial_{\mu\nu}\Phi_i)} \delta \Phi_i\biggr)\nonumber\\
&& + \biggl(\frac{\partial L}{\partial (\partial_{\mu\nu\lambda}\Phi_i)}\partial_{\nu\lambda}\delta\Phi_i
-\partial_\nu\frac{\partial L}{\partial (\partial_{\mu\nu\lambda}\Phi_i)}\partial_\lambda\delta\Phi_i
+\partial_{\nu\lambda}\frac{\partial L}{\partial (\partial_{\mu\nu\lambda}\Phi_i)}\delta\Phi_i\biggr)+\dots \nonumber \\
\label{eq.elc}
\eea
$\frac{\delta L}{\delta \Phi_i}$ has the role of $\hat{G}_\alpha$ and 
$j_\Phi^\mu$ has the role of $\mc{G}^\nu$.
$\frac{\delta L}{\delta \Phi_i}$ is the Euler--Lagrange derivative of $L$ with respect to $\Phi_i$, 
and the Euler--Lagrange equations
for $\Phi_i$ are $\frac{\delta L}{\delta \Phi_i}=0$. We call $j_\Phi^\mu$ the Euler--Lagrange current corresponding to $\Phi_i$,
because it is the counterpart of the Euler--Lagrange derivative in (\ref{eq.dl3}).
We note that a closely related quantity is called symplectic potential form in \cite{IW}.
Formulas similar to (\ref{eq.dl3}), (\ref{eq.el1}) and (\ref{eq.elc}) can also be
obtained for $\chi_j$
by applying (\ref{eq.g2}) to (\ref{eq.dl2}). 

Let us consider now a specific transformation 
and assume that $\chi_j$ are in a configuration that is invariant under this transformation, i.e.\ $\delta\chi_j=0$.
If in addition 
\beq
\label{eq.K}
\delta L_\Phi= \partial_\mu K^\mu 
\eeq
holds with some $K^\mu$,
then this transformation is called a symmetry transformation, 
and (\ref{eq.dl3}) implies that
\beq
\label{eq.n1}
\partial_\mu J_\Phi^\mu + \frac{\delta L}{\delta \Phi_i}\delta \Phi_i  = 0\ ,
\eeq
where $J_\Phi^\mu$ is defined as
\beq
\label{eq.n2}
J_\Phi^\mu = j_\Phi^\mu -K^\mu\ 
\eeq
and is called Noether current.
In particular if $\Phi_i$ satisfy their EL equations, then from (\ref{eq.n1}) it follows that
the current $J_\Phi^\mu$ is conserved: $\partial_\mu J_\Phi^\mu=0$.
These statements for $J_\Phi^\mu$ constitute the first theorem, extended to the situation when some of the elementary fields are fixed but their configuration is invariant under the symmetry that is considered.

Usually $\delta \Phi_i$ are local functions of $\Phi_i$, 
$K^\mu$ is also required to be a local function of $\Phi_i$, and (\ref{eq.K}) is understood to be an identity for $\Phi_i$. 
In the rest of the paper these properties will be assumed. 

Although (\ref{eq.K}) is required above to hold only for a specific configuration of $\chi_j$, 
in practice one often has an identity $\delta L = \partial_\mu K^\mu$ without any restriction on $\chi_j$ and
with $K^\mu$, $\delta\chi_j$, $\delta\Phi_i$ that are local functions of both $\Phi_i$ and $\chi_j$,
and then this identity reduces to (\ref{eq.K}) if $\chi_j$ is such that $\delta\chi_j=0$.

It is clear that $K^\mu$ is not uniquely determined in (\ref{eq.K}), 
therefore the application of the above theorem involves making a suitable choice. 
The simplest possibility, which is suitable for many cases, is $K^\mu=0$ (see Section \ref{sec.eymm} for examples).

The local conservation law $\partial_\mu J_\Phi^\mu=0$ can be rewritten in 
integral form by applying Stokes' theorem. Let $\mc{U}$ be a $D$-dimensional domain within $U$. By using Stokes' theorem one obtains 
\beq
\label{eq.intcons}
\int_{\partial \mc{U}} n_\mu J_\Phi^\mu = 0\ ,
\eeq
where $\partial \mc{U}$ is the boundary of $\mc{U}$
and $n_\mu$ is the normal vector field of $\partial \mc{U}$. $n_\mu$ is normalized using the flat Euclidean metric $\delta_{\mu\nu}$ determined by the coordinate system, i.e.\ $\delta^{\mu\nu}n_\mu n_\nu = 1$.
By restricting this metric to $\partial \mc{U}$ one gets a Riemannian metric on $\partial \mc{U}$, 
and the corresponding measure is the one that is used for the integration over $\partial \mc{U}$.
By choosing $\mc{U}$ to be a cylindrical domain $\mc{U}=[t_1,t_2]\times\Omega$, where $\Omega$ is a 
$D-1$ dimensional domain, one obtains from (\ref{eq.intcons}) the charge conservation law
\beq
\int_{\Omega}  J^0_\Phi|_{x^0=t_2} -  \int_{\Omega}  J^0_\Phi|_{x^0=t_1} = 
-\int _{[t_1,t_2]\times \partial\Omega} n_\mu J_\Phi^\mu \ ,
\eeq
where $\partial \Omega$ is the boundary of $\Omega$ and $n_\mu$ is the outward pointing
normal vector of  $\partial \Omega$. 
$\int_{\Omega}  J^0_\Phi|_{x^0=t_1}$ and $\int_{\Omega}  J^0_\Phi|_{x^0=t_2}$ are the charges in $\Omega$ 
at $x^0=t_1$ and $x^0=t_2$, respectively, and $\int _{[t_1,t_2]\times \partial\Omega} n_\mu J_\Phi^\mu$
is the charge that flows out of $\Omega$ during the $x^0$ interval $[t_1,t_2]$. The index of $x^\mu$ runs from $0$ to $D-1$.

\subsection{The second theorem}
\label{sec.n2}

Let us consider a transformation of the fields specified by $\delta \Phi_i$ and $\delta \chi_j$ of the form
\bea
\label{eq.t1}
\delta \Phi_i & = & \delta\Phi_{i\alpha}\, \epsilon^\alpha 
+\delta\Phi_{i\alpha}^\mu\, \partial_\mu\epsilon^\alpha
+\delta\Phi_{i\alpha}^{\mu\nu}\, \partial_{\mu\nu} \epsilon^\alpha +\dots\\
\label{eq.t2}
\delta \chi_j & = &  \delta\chi_{j\alpha}\, \epsilon^\alpha
+\delta\chi_{j\alpha}^\mu\, \partial_\mu\epsilon^\alpha
+\delta\chi_{j\alpha}^{\mu\nu}\, \partial_{\mu\nu} \epsilon^\alpha +\dots\ ,
\eea
where $\epsilon^\alpha$ is a function that can have several components indexed by $\alpha$
and may be commuting or anticommuting, 
and there are finitely many terms in the sums on the right hand sides. 
It is assumed that
$\delta\Phi_{i\alpha}$, $\delta\Phi_{i\alpha}^\mu$, $\delta\Phi_{i\alpha}^{\mu\nu}$, ..., 
$\delta\chi_{j\alpha}$, $\delta\chi_{j\alpha}^\mu$, $\delta\chi_{j\alpha}^{\mu\nu}$, ...
are local functions of $\Phi_i$ and $\chi_j$ and do not depend on $\epsilon^\alpha$ and on its derivatives.
Transformations of this form are called (infinitesimal) gauge transformations, parametrized by $\epsilon^\alpha$.
We do not assume any group property of these transformations.

The partial integration formula (\ref{eq.g2})
can be applied to $\frac{\delta L}{\delta\Phi_i}\delta\Phi_i$ and to $\frac{\delta L}{\delta\chi_j}\delta\chi_j$, giving
\bea
\label{eq.b1}
\frac{\delta L}{\delta\Phi_i}\delta\Phi_i & = & B_{\Phi\alpha} \epsilon^\alpha +\partial_\mu \mc{B}_\Phi^\mu\\
\label{eq.b2}
\frac{\delta L}{\delta\chi_j}\delta\chi_j & = &  B_{\chi\alpha}\epsilon^\alpha +\partial_\mu \mc{B}_\chi^\mu\ ,
\eea
where
\bea
\label{eq.bb1}
B_{\Phi\alpha} & = & \frac{\delta L}{\delta \Phi_i}\delta\Phi_{i\alpha}
-\partial_\mu \biggl(  \frac{\delta L}{\delta \Phi_i} \delta\Phi_{i\alpha}^\mu\biggr)
+ \partial_{\mu\nu} \biggl(\frac{\delta L}{\delta \Phi_i} \delta\Phi_{i\alpha}^{\mu\nu} \biggr) - \dots \\
\label{eq.bb2}
\mc{B}_\Phi^\mu & = &\frac{\delta L}{\delta \Phi_i}\delta\Phi_{i\alpha}^\mu  \epsilon^\alpha 
+\biggl[ \frac{\delta L}{\delta\Phi_i}\delta\Phi_{i\alpha}^{\mu\nu} \partial_\nu\epsilon^\alpha
- \partial_\nu \biggl(\frac{\delta L}{\delta\Phi_i}\delta\Phi_{i\alpha}^{\mu\nu} \biggr)\epsilon^\alpha\biggr] +\dots
\eea
and similar formulas can also be written for $\chi_j$. Using (\ref{eq.b1}) and (\ref{eq.b2}) one gets 
\bea
&&\hspace{-2cm}\delta L = \delta L_\Phi + \delta L_\chi =   \frac{\delta L}{\delta\Phi_i}\delta\Phi_i + \frac{\delta L}{\delta\chi_j}\delta\chi_j
+ \partial_\mu j_\Phi^\mu + \partial_\mu j_\chi^\mu \nonumber\\
&&=  B_{\Phi\alpha} \epsilon^\alpha +\partial_\mu \mc{B}_\Phi^\mu + \partial_\mu j_\Phi^\mu +
 B_{\chi\alpha}\epsilon^\alpha +\partial_\mu \mc{B}_\chi^\mu + \partial_\mu j_\chi^\mu\ .
\label{eq.b3}
\eea

Let us assume that the transformation specified by (\ref{eq.t1}) and (\ref{eq.t2}) is a symmetry for 
arbitrary $\epsilon^\alpha$ functions (and without any assumption on $\chi_j$),
i.e.\ 
\beq
\label{eq.b4}
\delta L = \partial_\mu K^\mu
\eeq
with some $K^\mu$. $K^\mu$ is also assumed to be a homogeneous linear local function of $\epsilon^\alpha$, 
i.e.
\beq
\label{eq.khom}
K^\mu = K^\mu_\alpha  \epsilon^\alpha +  K^{\mu\nu}_\alpha \partial_\nu \epsilon^\alpha + 
 K^{\mu\nu\lambda}_\alpha \partial_{\nu\lambda} \epsilon^\alpha + \dots\ ,
\eeq
where $K^\mu_\alpha$,  $K^{\mu\nu}_\alpha$, $K^{\mu\nu\lambda}_\alpha$, ... do not depend on $\epsilon^\alpha$ and on its derivatives 
and $K^{\mu\nu\lambda\dots}_\alpha$ are symmetric in the upper indices $\nu\lambda\dots$ following $\mu$. 
$K^\mu_\alpha$,  $K^{\mu\nu}_\alpha$, $K^{\mu\nu\lambda}_\alpha$, ...
are also local functions of $\Phi_i$ and $\chi_j$. In this case $L$ is said to have a gauge symmetry, and  
it follows from (\ref{eq.b3}) that
\beq
\label{eq.221}
 (B_{\Phi\alpha} + B_{\chi\alpha}) \epsilon^\alpha = -\partial_\mu (\mc{B}_\Phi^\mu + \mc{B}_\chi^\mu  +J^\mu)\ ,
\eeq
where 
\beq
\label{eq.J}
J^\mu=j_\Phi^\mu +j_\chi^\mu -K^\mu\ .
\eeq

If $\epsilon^\alpha$ and sufficiently many of its derivatives vanish on the boundary of an open domain $\Omega$, 
then by applying Stokes' theorem we get 
\beq
\label{eq.b5}
\int_{\Omega} d^D x\, (B_{\Phi\alpha} + B_{\chi\alpha})  \epsilon^\alpha = 0\ 
\eeq
from (\ref{eq.221}).
Since (\ref{eq.b4}) holds for arbitrary $\epsilon^\alpha$,
(\ref{eq.b5}) implies that  
\beq
\label{eq.bi}
B_{\Phi\alpha} + B_{\chi\alpha}=0\ .
\eeq
This result is the second theorem. 
From (\ref{eq.bb1})
it can be seen that (\ref{eq.bi}) is a differential identity for the EL derivatives of $L$ with respect to $\Phi_i$ and $\chi_j$.

(\ref{eq.bi}) applied to the Einstein--Hilbert Lagrangian with the diffeomorphism symmetry as the gauge symmetry
gives $\nabla_\mu G^{\mu\nu}=0$, where $G^{\mu\nu}$ is the 
Einstein tensor. This is the twice contracted Bianchi identity, therefore (\ref{eq.bi}) can be called generalized Bianchi identity.
$B_{\Phi\alpha}$ and $B_{\chi\alpha}$ will be called Bianchi expressions.

Since $\mc{B}_\Phi^\mu$ and $\mc{B}_\chi^\mu$ are the counterparts of $B_{\Phi\alpha}$ and $B_{\chi\alpha}$ in (\ref{eq.b1}) and (\ref{eq.b2}), we call them Bianchi currents. 
Note that $\mc{B}_\Phi^\mu=0$ if $\Phi_i$ satisfy their EL equations and obviously the same is true for $\chi_j$ and  $\mc{B}_\chi^\mu$.  
Moreover,  $\mc{B}_\Phi^\mu=0$ also if the coefficients of all derivatives of $\epsilon^\alpha$ 
on the right hand side of (\ref{eq.t1}) are zero, 
and the same can be said of $\mc{B}_\chi^\mu$. 

If $L$ has a gauge symmetry, then the corresponding variation 
$\delta L$ of $L$ is also a homogeneous linear local function of $\epsilon^\alpha$ with some coefficients
$\delta L_\alpha$, $\delta L_\alpha^\mu$, $\delta L_\alpha^{\mu\nu}$, ... 
and (\ref{eq.b4}) holds for arbitrary $\epsilon^\alpha$, 
therefore the coefficients of $\epsilon^\alpha$, $\partial_\mu \epsilon^\alpha$, $\partial_{\mu\nu}\epsilon^\alpha$, ... 
on the two sides of (\ref{eq.b4})
have to be equal, i.e.\
\bea
\label{eq.kn1}
\delta L_\alpha & = & \partial_\mu K_\alpha^\mu \\
\label{eq.kn2}
\delta L_\alpha^\mu & = &  K_\alpha^\mu + \partial_\nu K_\alpha^{\nu\mu} \\
\label{eq.kn3}
\delta L_\alpha^{\mu\nu} & = & K_\alpha^{\mu\nu} + \partial_\lambda K_\alpha^{\lambda\mu\nu} \\
\dots && \nonumber
\eea
These equations are known as Klein--Noether identities.

\subsection{The third theorem}
\label{sec.n3}

In this section we continue to consider Lagrangian systems with gauge symmetry.
The current $J^\mu$ introduced in (\ref{eq.J}) is the standard Noether current corresponding to the
symmetry transformation (\ref{eq.t1}), (\ref{eq.t2}) in absence of any fixed field, thus it 
is conserved if both $\Phi_i$ and $\chi_j$ satisfy their EL equations. 
On the other hand, from (\ref{eq.221}) and (\ref{eq.bi}) it follows that
the current 
\beq
\label{eq.I}
I^\mu = \mc{B}_\Phi^\mu + \mc{B}_\chi^\mu  + J^\mu
\eeq
is also conserved, regardless of the EL equations. 
Nevertheless, if $\Phi_i$ and $\chi_j$ satisfy their EL equations, then $I^\mu = J^\mu$.
Thus, by adding $\mc{B}_\Phi^\mu + \mc{B}_\chi^\mu$ to $J^\mu$ we get a current which is conserved regardless of the 
EL equations, but which nevertheless coincides with $J^\mu$ if the 
EL equations of all fields are satisfied. 
This is the first part of the third theorem. 

The second part is the following: since $I^\mu$ is a homogeneous linear local function of $\epsilon^\alpha$
and is conserved for arbitrary $\epsilon^\alpha$, it can be written as
\beq
\label{eq.i1}
I^\mu = \partial_\nu \Sigma^{\mu\nu}\ ,
\eeq
where $\Sigma^{\mu\nu}$, which is called superpotential, is antisymmetric and is given by the formula
\bea
&&\hspace{-8mm}\Sigma^{\mu\nu} =   \frac{1}{2} (I_\alpha^{\mu\nu} -  I_\alpha^{\nu\mu}) \epsilon^\alpha
+\biggl[\frac{2}{3} (I_\alpha^{\mu\nu\rho}-I_\alpha^{\nu\mu\rho})\partial_\rho\epsilon^\alpha
-\frac{1}{3}\partial_\rho (I_\alpha^{\mu\nu\rho}- I_\alpha^{\nu\mu\rho})\epsilon^\alpha\biggr]\nonumber\\
&&\hspace{-8mm}+\biggl[\frac{3}{4} (I_\alpha^{\mu\nu\rho\lambda}-I_\alpha^{\nu\mu\rho\lambda})\partial_{\rho\lambda}\epsilon^\alpha
-\frac{2}{4}\partial_\lambda (I_\alpha^{\mu\nu\rho\lambda}-I_\alpha^{\nu\mu\rho\lambda})\partial_\rho\epsilon^\alpha
+\frac{1}{4} \partial_{\rho\lambda}(I_\alpha^{\mu\nu\rho\lambda}-I_\alpha^{\nu\mu\rho\lambda})\epsilon^\alpha
\biggr]+\dots\nonumber\\
\label{eq.i2}
\eea
with the $n$-th group of terms on the right hand side being
\beq
\label{eq.i3}
\sum_{i=1}^{n}(-1)^{i+1}\frac{n+1-i}{n+1}
\partial_{\lambda_{n-i+1}\dots\lambda_{n-1}}(I^{\mu\nu\lambda_1\dots\lambda_{n-1}}_\alpha - I^{\nu\mu\lambda_1\dots\lambda_{n-1}}_\alpha)  \partial_{\lambda_1\lambda_2\dots \lambda_{n-i}}\epsilon^\alpha \ .
\eeq
The $I_\alpha^{\mu\nu\lambda\dots}$ appearing in these formulas are the coefficients in the expansion
\beq
\label{eq.i4}
I^\mu = I^\mu_\alpha  \epsilon^\alpha +  I_\alpha^{\mu\nu} \partial_\nu \epsilon^\alpha + 
 I_\alpha^{\mu\nu\lambda} \partial_{\nu\lambda} \epsilon^\alpha + \dots \ .
\eeq
$I_\alpha^{\mu\nu\lambda\dots}$ are symmetric in the upper indices $\nu\lambda\dots$ following $\mu$.
The two conditions mentioned above (\ref{eq.i1}) are sufficient for (\ref{eq.i1}) and (\ref{eq.i2}), i.e.\ 
no more details on the properties of $I^\mu$ are needed. 
This second part of the third theorem is proved for example in \cite{Fletcher,WaldJMP}. 

It should be noted that although from the Poincar\'e lemma it follows that in any simply connected domain there exists some antisymmetric $\Sigma^{\mu\nu}$
for which (\ref{eq.i1}) holds, the statement that the $\Sigma^{\mu\nu}$ given by  (\ref{eq.i2}) satisfies (\ref{eq.i1}) is stronger. (\ref{eq.i2}) implies that $\Sigma^{\mu\nu}$ is also a local function of the fields, and therefore by applying Stokes' theorem the total charge corresponding to $I^\mu$ in a $D-1$ dimensional domain $\Omega$ can be expressed as an integral of a local expression of the fields over the boundary of $\Omega$.  

Currents of the form $\partial_\nu \Sigma^{\mu\nu}$, where $\Sigma^{\mu\nu}$ is antisymmetric, are identically conserved 
because of the antisymmetry of $\Sigma^{\mu\nu}$. Moreover, from Stokes' theorem it follows that if $\Sigma^{\mu\nu}$ falls off sufficiently fast at infinity, then the total charge associated with the current $\partial_\nu \Sigma^{\mu\nu}$ is zero. 
Due to these facts the currents that have the form $\partial_\nu \Sigma^{\mu\nu}$ are often called trivial. 
Nevertheless, the $\Sigma^{\mu\nu}$ introduced above does not always fall off very fast at infinity and the 
associated charge can be different from zero. 
For further discussion of the definition of total charges in gauge theories, especially in general relativity, the reader is referred to the literature, e.g.\ \cite{WaldGR,IW,Szabados,BB,PL,PL2,Petrov,WZ,JM,OR,Prabhu,HS,Hajian,MTW}.

\subsection{Conserved currents in gauge theories in the presence of fixed fields}
\label{sec.n4}

In this section the generalization of the construction mentioned at the beginning of Section \ref{sec.intr} is described.
It is assumed that $L$ has a gauge symmetry in the general sense described in Section \ref{sec.n2}.
This means that $L$ is assumed to have a gauge symmetry with respect to the complete set of fields, 
but with respect to the actual dynamical fields it does not necessarily have any gauge symmetry.

If $\Phi_i$ satisfy their EL equations, then $B_{\Phi\alpha}=0$, as can be seen from
(\ref{eq.bb1}), and thus from the generalized Bianchi identity (\ref{eq.bi}) it follows that 
\beq
\label{eq.bi2}
B_{\chi\alpha}=0\ ,
\eeq
without any assumption on $\chi_j$. This is the first part of the construction. 
We call (\ref{eq.bi2}) a partial Bianchi identity.

Next, let us consider the situation in which $\delta\chi_j=0$ 
for some particular $\epsilon^\alpha$, 
i.e.\ the configuration of $\chi_j$ is invariant with respect to a particular infinitesimal gauge transformation. 
In this case it follows from (\ref{eq.b2}) that
\beq
\label{eq.45}
0=  B_{\chi\alpha} \epsilon^\alpha + \partial_\mu \mc{B}_\chi^\mu\ .
\eeq
Moreover, if $\Phi_i$ satisfy their EL equations, 
then $B_{\chi\alpha}=0$, as we have seen, thus (\ref{eq.45}) reduces to
\beq
\label{eq.43}
0= \partial_\mu  \mc{B}_\chi^\mu\ . 
\eeq
This means that in addition to the standard Noether current $J_\Phi^\mu$ there is another conserved current, $\mc{B}_\chi^\mu$.
This is the second part of the construction.

If $\delta\chi_j=0$, then  
$j_\chi^\mu=0$ and thus $J^\mu=J^\mu_\Phi$. 
From (\ref{eq.I}) and (\ref{eq.i1}) it follows then that
\beq
\label{eq.44}
-\mc{B}_\chi^\mu = J_\Phi^\mu + \mc{B}_\Phi^\mu - I^\mu = J_\Phi^\mu + \mc{B}_\Phi^\mu -  \partial_\nu \Sigma^{\mu\nu}\ ,
\eeq
therefore the difference of $-\mc{B}_\chi^\mu$ and $J_\Phi^\mu$ is the sum of the terms
$\mc{B}_\Phi^\mu$ and $-\partial_\nu \Sigma^{\mu\nu}$,
the first of which vanishes when $\Phi_i$ satisfy their EL equations, 
and the second of which is the divergence of an antisymmetric matrix. This is the third part of the construction.

(\ref{eq.44}) shows that although $J_\Phi^\mu$ and $\mc{B}_\chi^\mu$ are defined completely differently, 
they are equivalent in the usual sense of the equivalence of conserved currents. 
It is also worth mentioning
that $J_\Phi^\mu$ depends on the choice of $K^\mu$, whereas $\mc{B}_\chi^\mu$ does not. 

Until now it has been assumed that $L$ is the total Lagrangian of the system under consideration, 
but it is clear that if $L$ takes the form $L=L_1+L_2$, where $L_1$ does not depend on $\Phi_i$ and their derivatives 
and $L_2$ has gauge symmetry in itself, then the above construction can also be applied to $L_2$. 
For example, in the standard case in general relativity one constructs $\sqrt{-g}\,T^{\mu\nu}h_\nu$ from the matter Lagrangian rather than 
from the total Lagrangian, and the same will be done in the examples in Sections \ref{sec.eymm} and \ref{sec.5} as well.

If $L_1$ has a gauge symmetry in itself and $\chi_j$ satisfy their EL equations $\frac{\delta L_1}{\delta \chi_j}+\frac{\delta L_2}{\delta \chi_j} =0$, then
the generalized Bianchi identity for $L_1$ together with the EL equations 
of $\chi_j$ imply an equation  
that formally coincides with the 
partial Bianchi identity (\ref{eq.bi2}) obtained from $L_2$. 
In order to see this, one should consider that the expression that appears on the left hand side of the generalized Bianchi identity is given in the present case by the   
formula (\ref{eq.bb1}) for Bianchi expressions, with $L\to L_1$ and $\Phi\to \chi$. 
In this expression $\frac{\delta L_1}{\delta \chi_j}$ can be replaced by 
$-\frac{\delta L_2}{\delta \chi_j}$ as a consequence of the EL equations of $\chi_j$, and the resulting 
expression is just $-1$ times the one that appears in (\ref{eq.bi2}), 
thus indeed (\ref{eq.bi2}) is obtained.  
In this argument it is not necessary to use any symmetry property of $L_2$ and $\Phi_i$ do not have to satisfy their EL equations. 
The example in Section \ref{sec.eymm} shows that the observations 
that the Einstein equation implies the divergencelessness of the energy-momentum tensor 
and the Maxwell equation implies the conservation of the electric current
are special cases of the result 
described in this paragraph.

We note that 
the second part of the construction is not used in the derivation of (\ref{eq.44}), and (\ref{eq.44}) also implies the conservation of $\mc{B}_\chi^\mu$. On the other hand, it is the derivation given in the second part that is the generalization of the usual derivation of the conservation of $\sqrt{-g}\,T^{\mu\nu}h_\nu$ in general relativity. 

The results in Section $2$ of \cite{Borokhov} are similar to (\ref{eq.44}),
but they are less explicit and are restricted
to the special case when the particular infinitesimal gauge transformation under which $\chi_j$ are invariant is rigid, i.e.\ 
is such that $\epsilon^\alpha$ is a constant function. 
In principle, it is not a severe restriction to consider only rigid transformations, 
since a non-rigid gauge transformation can be transformed into a rigid one by suitable reparametrisation. 
For example, let us choose $n$ different functions $\epsilon^\alpha_k$, $k=1,\dots,n$, so that
$\epsilon^\alpha_1$ corresponds to the particular transformation under which $\delta \chi_j=0$, and introduce the new parameter functions $\xi^k$ via $\epsilon^\alpha = \xi^k \epsilon^\alpha_k$. 
In terms of $\xi^k$ the transformation specified by $\epsilon^\alpha=\epsilon^\alpha_1$ is rigid, since it
corresponds to $\xi^1=1$, $\xi^k=0$, $k=2,\dots,n$. 
Nevertheless, one can deal with non-rigid transformations directly, as we have seen in this section.

\section{Matter fields in the presence of fixed gravitational and Yang--Mills gauge fields}
\label{sec.eymm}

In this section the case of matter fields in the presence of gravitational and YM gauge fields is discussed. 
Since fermionic fields are also considered, the basic gravitational field variable is taken to be an orthonormal tetrad field $V_\mu^{\bar{\mu}}$.
The metric can be expressed in terms of $V_\mu^{\bar{\mu}}$ as $g_{\mu\nu}=V_\mu^{\bar{\mu}} V_\nu^{\bar{\nu}}g_{\bar{\mu}\bar{\nu}}$, where $g_{\bar{\mu}\bar{\nu}}=\mathrm{diag}(1,-1,-1,-1)$. Here and in the following an overbar is used to distinguish internal Lorentz vector indices. Such indices can be raised and lowered by $g_{\bar{\mu}\bar{\nu}}$ and its inverse and $V_\mu^{\bar{\mu}}$ and its inverse can be used to turn spacetime vector indices into internal Lorentz vector indices and vice versa.
For a detailed introduction to the tetrad formalism and its use for including half integer spin fields in general relativity, 
see e.g.\ \cite{W}. A different formalism is described e.g.\ in \cite{GL}.

The Lie algebra of the global YM gauge group is taken to be a direct sum of compact simple and $u(1)$ algebras.
It would be straightforward to consider several YM gauge groups with distinct coupling constants, 
but for simplicity only one is taken.
The structure constants of the Lie algebra of the global YM gauge group are denoted by $f\indices{_{ab}^c}$. 
The basis of the Lie algebra is chosen so that $\delta^{ab}$ 
is invariant under the adjoint action, 
therefore there is no significant difference between upper and lower Lie algebra indices. 
Nevertheless, upper and lower indices will be distinguished
and $\delta^{ab}$ will be used to raise and lower them. $f^{abc}$ is completely antisymmetric and real.
The YM field strength is defined as
$F_{\mu\nu}^a = \partial_\mu A_\nu^a - \partial_\nu A_\mu^a + \kappa f\indices{^a_{bc}}A_\mu^b A_\nu^c$, 
where $\kappa$ is the YM coupling constant and $A_\mu^a$ is the YM vector potential. 

The Lagrangian density function $L$ of the matter fields is assumed to be
an even real local function of the tetrad field, the YM vector potential and the matter fields,
and it is assumed to have diffeomorphism, local Lorentz and YM gauge symmetry in the sense described in Section \ref{sec.n2}. 
The precise form of $L$ and of the Lagrangian of the gravitational and 
YM gauge fields does not need to be specified. For fermionic fields Lorentz transformations mean $SL(2,\CC)$ transformations.

The application of the first two parts of the construction 
described in Section \ref{sec.n4} is discussed in Section \ref{sec.app1}, 
and the third part is discussed in Section \ref{sec.app2}. 
The matter fields will take the role of $\Phi_i$, 
whereas the tetrad and YM gauge fields together will take the role of $\chi_j$. 
As was mentioned in Section \ref{sec.n4}, the construction will be applied to the matter Lagrangian described above, 
rather than to the total Lagrangian. 
The total gauge symmetry group that will be considered is the group generated by the symmetries mentioned above.

\subsection{Bianchi currents and partial Bianchi identities}
\label{sec.app1}

For the application of the first two parts of the construction described in Section \ref{sec.n4} 
it is necessary to know the transformation properties of the tetrad and of the YM gauge fields.
The $K^\mu$ quantities in the symmetry condition (\ref{eq.b4}) and the transformation properties of the matter fields are not needed. 

About the Bianchi currents the following general preliminary remarks can be made: since the first order variations
of each field that we consider depend only on $\epsilon^\alpha$ and $\partial_\mu\epsilon^\alpha$
and not on higher derivatives, the corresponding Bianchi currents do not depend on the derivatives of $\epsilon^\alpha$ at all. 
Specifically, the first term on the right hand side of (\ref{eq.bb2}) gives the Bianchi currents. 
Furthermore, if the first order variation of a field does not depend on $\partial_\mu\epsilon^\alpha$, 
then the corresponding Bianchi current is zero.

The first order variation of the tetrad and the YM vector potential under a YM gauge transformation
parametrized by $\epsilon^a$, which takes the role of $\epsilon^\alpha$
and has values in the coadjoint representation of the global gauge group, is 
\bea
\label{eq.ym1}
\delta V_\mu^{\bar{\mu}} & = & 0 \\
\label{eq.ym2}
\delta A_\mu^a & = & D_\mu\epsilon^a\ . 
\eea 
$D_\mu$ denotes the covariant differential operator for the whole gauge group---for more detail on its definition see Appendix \ref{sec.A}. 
In (\ref{eq.ym2}) $D_\mu \epsilon^a = \partial_\mu \epsilon^a + \kappa f\indices{^a_{bc}}A_\mu^b \epsilon^c$. 

In the case of local Lorentz transformations the role of $\epsilon^\alpha$ is taken by $\omega^{\bar{\mu}\bar{\nu}}$, 
which is antisymmetric. The first order variation of the tetrad and the YM vector potential is
\bea
\label{eq.ll1}
\delta V_\mu^{\bar{\mu}} & = & -\omega^{\bar{\lambda}\bar{\rho}}(L_{\bar{\lambda}\bar{\rho}})\indices{_{\bar{\nu}}^{\bar{\mu}}}
V_\mu^{\bar{\nu}} \\
\label{eq.ll2}
\delta A_\mu^a & = & 0\ . 
\eea 
The $(L_{\bar{\lambda}\bar{\rho}})\indices{_{\bar{\nu}}^{\bar{\mu}}}=
g_{\bar{\lambda}\bar{\nu}}\delta_{\bar{\rho}}^{\bar{\mu}}-g_{\bar{\rho}\bar{\nu}}\delta_{\bar{\lambda}}^{\bar{\mu}}$ 
appearing in (\ref{eq.ll1}) are the generators of the 
Lorentz group in the Minkowski representation.

In the case of diffeomorphisms the role of $\epsilon^\alpha$ is taken by the vector fields $h^\mu$, which generate  
diffeomorphisms. The first order variation of the tetrad and the YM vector potential is 
\bea
\label{eq.diff1}
\delta V_\mu^{\bar{\mu}} & = & -h^\nu\partial_\nu V_\mu^{\bar{\mu}} - \partial_\mu h^\nu  V_\nu^{\bar{\mu}} \\
\label{eq.diff2}
\delta A_\mu^a & = & -h^\nu \partial_\nu A_\mu^a - \partial_\mu h^\nu A_\nu^a\ . 
\eea
(\ref{eq.diff1}) implies that $\delta g_{\mu\nu} = 
-h^\lambda\partial_\lambda g_{\mu\nu} - g_{\nu\lambda}\partial_\mu h^\lambda - g_{\mu\lambda}\partial_\nu h^\lambda = 
-\nabla_\mu h_\nu -\nabla_\nu h_\mu$. 
Here and in subsequent formulas $\nabla_\mu$ denotes the usual Levi--Civita covariant differential operator associated with the metric 
$g_{\mu\nu}$. $\nabla_\mu$ acts only on the spacetime vector and covector indices. 
(\ref{eq.diff1}) and (\ref{eq.diff2}) can also be written as 
$\delta V_\mu^{\bar{\mu}}  =  -h^\nu\nabla_\nu V_\mu^{\bar{\mu}} - \nabla_\mu h^\nu  V_\nu^{\bar{\mu}}$, 
$\delta A_\mu^a  =  -h^\nu \nabla_\nu A_\mu^a - \nabla_\mu h^\nu A_\nu^a$.

Using these transformation properties one finds that 
in the case of the YM gauge symmetry the partial Bianchi identity (\ref{eq.bi2}) takes the form
\beq
\label{eq.ymb1}
B_{Aa}=\sqrt{-g}D_\mu \mc{J}^\mu_a = 0\ ,
\eeq
where $\mc{J}^\mu_a$ is defined as
\beq
\label{eq.ymb2}
\sqrt{-g} \mc{J}^\mu_a = -\frac{\delta L}{\delta A_\mu^a}\ 
\eeq
and $g$ denotes the determinant of the metric.
$B_{Va}$ is obviously $0$ in this case. We note that (\ref{eq.ymb1}) was also obtained e.g.\ in \cite{BBGS}.
As is well known, (\ref{eq.ymb1}) also follows from the YM equation, if the YM field is not fixed. 
The Bianchi currents $\mc{B}_A^\mu$ and $\mc{B}_V^\mu$, corresponding to the YM gauge field and to the tetrad, are found to be 
\beq
\label{eq.ymb3a}
\mc{B}_A^\mu = -\sqrt{-g}\, \epsilon^a \mc{J}^\mu_a\ ,
\qquad \mc{B}_V^\mu=0\ ,
\eeq
thus $\mc{B}_{\chi}^\mu = \mc{B}_A^\mu + \mc{B}_V^\mu$ is 
\beq
\label{eq.ymb3b}
\mc{B}_{\chi}^\mu = -\sqrt{-g}\, \epsilon^a \mc{J}^\mu_a\ .
\eeq
$\mc{B}_{\chi}^\mu$ is conserved if the first order variation of $A_\mu^a$ corresponding to $\epsilon^a$ is zero, 
i.e.\ if $D_\mu \epsilon^a = 0$, and the matter fields satisfy their EL equations. It is easy to see that this is a consequence of the partial Bianchi identity (\ref{eq.ymb1}):
$\partial_\mu \mc{B}_{\chi}^\mu = -\sqrt{-g}\, \nabla_\mu (\epsilon^a \mc{J}^\mu_a)
=  -\sqrt{-g} (D_\mu \epsilon^a \mc{J}^\mu_a + \epsilon^a D_\mu \mc{J}^\mu_a)$, thus in virtue of (\ref{eq.ymb1})
$\partial_\mu \mc{B}_{\chi}^\mu =-\sqrt{-g} D_\mu \epsilon^a \mc{J}^\mu_a$, and this is zero if $D_\mu \epsilon^a = 0$.
This derivation is a specialization of the general derivation of (\ref{eq.43}) in Section \ref{sec.n4}.
For flat spacetime and in slightly different context it can also be found in \cite{AD}.

In the case of the local Lorentz symmetry the partial Bianchi identity (\ref{eq.bi2}) takes the form 
\beq
\label{eq.lb1}
B_{\bar{\lambda}\bar{\rho}}= B_{V,\bar{\lambda}\bar{\rho}} =   -2\sqrt{-g}\,\mathbb{T}_{\bar{\lambda}\bar{\rho}} = 0
\eeq
where $\mathbb{T}_{\bar{\lambda}\bar{\rho}}$ is the antisymmetric part in the decomposition 
\beq
\label{eq.lb2}
V_{\bar{\lambda}\rho}\frac{\delta L}{\delta V_\rho^{\bar{\rho}}}=\sqrt{-g}(-T\indices{_{\bar{\lambda}\bar{\rho}}} + \mathbb{T}\indices{_{\bar{\lambda}\bar{\rho}}})
\eeq
of $\frac{1}{\sqrt{-g}}V_{\bar{\lambda}\rho}\frac{\delta L}{\delta V_\rho^{\bar{\rho}}}$ into symmetric and antisymmetric parts.
The Bianchi currents $\mc{B}_{V}^\mu$ and $\mc{B}_{A}^\mu$ are obviously $0$ for local Lorentz symmetry,
thus also $\mc{B}_{\chi}^\mu = 0$. 

It is well known that (\ref{eq.lb1}) holds if all fields except the tetrad satisfy their EL equations \cite{W}, 
but, as we have seen, due to $\delta A_\mu^a=0$ (\ref{eq.lb1}) holds even when the YM vector potential is also fixed. 
It is clear that  
(\ref{eq.lb1}) holds generally if the fixed fields, with the exception of the tetrad, 
are scalar under local Lorentz transformations.

For the diffeomorphism symmetry the Bianchi expressions $B_{V\mu}$ and $B_{A\mu}$ are
\beq
\label{eq.bvmu}
B_{V\mu} =  \sqrt{-g}(-\nabla_\nu T\indices{^\nu_\mu}
+\nabla_\nu \mathbb{T}\indices{^\nu_\mu}
+\mathbb{T}^{\lambda\nu}V_{\lambda\bar{\mu}}\nabla_\mu V_\nu^{\bar{\mu}})
\eeq
\beq
\label{eq.bamu}
B_{A\mu} =  \sqrt{-g}(-\mc{J}^\nu_aF^a_{\nu\mu}
-D_\nu \mc{J}^\nu_a A_\mu^a)\ .
\eeq
The partial Bianchi identity (\ref{eq.bi2}) is $B_{V\mu} + B_{A\mu}=0$, which can nevertheless be simplified to the form  
\beq
\label{eq.db1}
B_\mu= \sqrt{-g}(-\nabla_\nu T\indices{^\nu_\mu}
-\mc{J}^\nu_a F^a_{\nu\mu}) = 0
\eeq
by using (\ref{eq.ymb1}) and (\ref{eq.lb1}).
$T^{\nu\mu}$, defined in (\ref{eq.lb2}), is the Einstein--Hilbert energy-momentum tensor.

The Bianchi currents $\mc{B}_V^\mu$ and  $\mc{B}_A^\mu$ are
\beq
\label{eq.bva}
\mc{B}_V^\mu =  \sqrt{-g}(T\indices{^\mu_\nu}
-\mathbb{T}\indices{^\mu_\nu})h^\nu\ ,
\qquad \mc{B}_A^\mu =  \sqrt{-g}\, h^\nu A_\nu^a \mc{J}^\mu_a\ .
\eeq
If the matter fields satisfy their EL equations, 
then $\mc{B}_{\chi}^\mu = \mc{B}_V^\mu + \mc{B}_A^\mu$ can be simplified to the form
\beq
\label{eq.db2}
\mc{B}^\mu =  \sqrt{-g}({T}\indices{^\mu_\nu} h^\nu+
h^\nu A_\nu^a \mc{J}^\mu_a)
\eeq
by using (\ref{eq.lb1}).
$\mc{B}^\mu$ is conserved if the matter fields satisfy their EL equations, $h^\mu$ is a Killing vector field and $A_\mu^a$ is also invariant under the diffeomorphisms 
generated by $h^\mu$. It should be noted that the invariance of the tetrad field is not necessary for the conservation of $\mc{B}^\mu$, 
the invariance of the metric (and of the vector potential) is sufficient. 
The reason for this is explained in the paragraph below (\ref{eq.dyb}). 

The conservation of $\mc{B}^\mu$ can again be seen as a consequence of (\ref{eq.db1}) and of the 
symmetry properties of the metric and the vector potential. 
(\ref{eq.ymb1}) is also needed, because it was also used to bring $B_\mu$ to the form (\ref{eq.db1}).
The first step of the derivation is to take the divergence of $\mc{B}^\mu$:  
$\partial_\mu \mc{B}^\mu = \sqrt{-g}(\nabla_\mu T\indices{^\mu_\nu} h^\nu + 
T\indices{^\mu_\nu} \nabla_\mu h^\nu + D_\mu \mc{J}^\mu_a A_\nu^a h^\nu 
+  \mc{J}^\mu_a D_\mu  A_\nu^a h^\nu
+ \mc{J}^\mu_a A_\nu^a \nabla_\mu h^\nu)$.
If $h^\mu$ is a Killing vector field, then $T\indices{^\mu_\nu} \nabla_\mu h^\nu = 0$. 
In virtue of the partial Bianchi identity (\ref{eq.ymb1}), also $D_\mu \mc{J}^\mu_a A_\nu^a h^\nu = 0$. 
According to (\ref{eq.db1}), $\nabla_\mu T\indices{^\mu_\nu} h^\nu$ can be replaced by 
$-\mc{J}^\mu_a F_{\mu\nu}^a h^\nu$. Thus the expression for $\partial_\mu \mc{B}^\mu$ can be written as
$\sqrt{-g} ( -\mc{J}^\mu_a F_{\mu\nu}^a h^\nu +
\mc{J}^\mu_a D_\mu  A_\nu^a h^\nu +
\mc{J}^\mu_a A_\nu^a \nabla_\mu h^\nu)$. 
Using the equation $h^\mu \nabla_\mu A_\nu^a + \nabla_\nu h^\mu A_\mu^a=0$, 
which expresses the invariance of the vector potential, 
$\mc{J}^\mu_a A_\nu^a \nabla_\mu h^\nu$ can be replaced by 
$-\mc{J}^\mu_a \nabla_\nu A_\mu^a  h^\nu$, thus
$\partial_\mu \mc{B}^\mu = 
\sqrt{-g} ( -\mc{J}^\mu_a F_{\mu\nu}^a h^\nu +
\mc{J}^\mu_a D_\mu  A_\nu^a h^\nu 
-\mc{J}^\mu_a \nabla_\nu A_\mu^a  h^\nu )$. 
This is clearly zero, since 
$F_{\mu\nu}^a = D_\mu A_\nu ^a - \nabla_\nu A_\mu^a$.

More generally, it is interesting to consider the situation when $h^\mu$ is a Killing vector field but the vector potential 
is invariant under the generated diffeomorphisms only up to a YM gauge transformation (see e.g.\ \cite{FM} regarding such invariance conditions). This means that there exists a 
gauge transformation parameter $\epsilon^a$ so that the vector potential is invariant under 
the joint infinitesimal diffeomorphism and YM gauge transformation corresponding to $(h^\mu,\epsilon^a)$: 
\beq
\label{eq.ainv}
\delta A_\nu^a = - h^\mu\partial_\mu A_\nu^a - \partial_\nu h^\mu A_\mu^a + D_\nu \epsilon^a = 0\ .
\eeq
The corresponding Bianchi current 
\beq
\label{eq.dyb}
\mc{B}^\mu =  \sqrt{-g}\,[{T}\indices{^\mu_\nu}
h^\nu+
(h^\nu A_\nu^a -\epsilon^a) \mc{J}^\mu_a]
\eeq
is the combination of (\ref{eq.ymb3b}) and (\ref{eq.db2}). The conservation of (\ref{eq.dyb}) can be derived in a similar way,
using (\ref{eq.ainv}), as the conservation of (\ref{eq.ymb3b}) and (\ref{eq.db2}). In this derivation the equivalent form
$\delta A_\nu^a = - h^\mu\nabla_\mu A_\nu^a - \nabla_\nu h^\mu A_\mu^a + D_\nu \epsilon^a = 0$ of (\ref{eq.ainv}) is useful.

In principle, the cases when the tetrad field is invariant under the diffeomorphisms generated by $h^\mu$ 
and when it is invariant only up to local Lorentz transformations are also different, but since the Bianchi current
corresponding to local Lorentz transformations is zero, the Bianchi currents (\ref{eq.db2}) or (\ref{eq.dyb})
are the same in both cases. This is not true for the Noether currents; see Sections \ref{sec.ll} and \ref{sec.mixedsymm}.

In the case when the YM gauge field is the electromagnetic field
the Bianchi current (\ref{eq.db2}) is precisely the current (\ref{eq.0}) proposed in \cite{NQV}, 
and the partial Bianchi identity (\ref{eq.db1}) is the Lorentz law (\ref{eq.1}). 
It should be noted that in the (charged black hole) configurations considered in \cite{NQV} the background electromagnetic field 
satisfies the Maxwell equation $\nabla_\mu F^{\mu\nu}=0$, and this property was used in \cite{NQV} 
in the derivation of the Lorentz law, but in the present derivation this is not needed.

Making a small digression, we mention that 
if the YM symmetry of the Lagrangian is not required and $A_\mu^a$ are merely some fixed covector fields, 
then $D_\nu \mc{J}^\nu_a=0$ does not necessarily hold, and
from (\ref{eq.bvmu}), (\ref{eq.bamu}) and (\ref{eq.lb1}) one obtains the partial Bianchi identity 
\beq
\label{eq.db3}
B_\mu= \sqrt{-g}(-\nabla_\nu T\indices{^\nu_\mu}
-\mc{J}^\nu_a F^a_{\nu\mu} - D_\nu \mc{J}^\nu_a A_\mu^a  ) = 0\ ,
\eeq
where $D_\mu$ and $F^a_{\nu\mu}$ are defined in the same way as in abelian gauge theory.
This is the identity that appears e.g.\ in \cite{DM1,DM2}.  
The formula (\ref{eq.db2}) for the Bianchi current for a diffeomorphism symmetry of the metric and $A_\mu^a$ remains unchanged.

\subsection{Noether currents and superpotentials}
\label{sec.app2}

In this section we discuss the Noether currents and the superpotentials for the diffeomorphism, 
local Lorentz and YM gauge symmetries, concentrating on the third part of the construction in Section \ref{sec.n4}.

For simplicity we assume that the Lagrangian density of the matter fields
takes the form
\beq
\label{eq.Lm}
L=\sqrt{-g}\,\hat{L}\bigl(V_\mu^{\bar{\mu}},\psi\indices{_{\check{\nu}}^{\check{\rho}}_{\alpha k}},
\bar{\psi}\indices{_{\check{\nu}}^{\check{\rho}\alpha k}},
D_\mu \psi\indices{_{\check{\nu}}^{\check{\rho}}_{\alpha k}},
D_\mu \bar{\psi}\indices{_{\check{\nu}}^{\check{\rho}\alpha k}}\bigr)\ ,
\eeq
where $\psi\indices{_{\check{\nu}}^{\check{\rho}}_{\alpha k}}$
is a complex matter field and $\bar{\psi}\indices{_{\check{\nu}}^{\check{\rho}\alpha k}}$ is its (generalized) Dirac conjugate. 
We take only one matter field, but it is completely straightforward to include 
more of them.  
The above form of $L$ implies 
that the matter field is coupled minimally to the gravitational and YM fields, and that $L$
does not depend on higher than first derivatives of the matter field. Matter fields can be real as well, 
in this case the Dirac conjugate field is not needed. 
$\psi\indices{_{\check{\nu}}^{\check{\rho}}_{\alpha k}}$ has $n$ covector indices and $m$ vector indices, 
denoted collectively as $\check{\nu}$ and $\check{\rho}$, i.e.\ 
$\check{\nu}\equiv \nu_1\nu_2\dots\nu_n$ and $\check{\rho}\equiv \rho_1\rho_2\dots\rho_m$.

Under the global YM gauge group $\psi\indices{_{\check{\nu}}^{\check{\rho}}_{\alpha k}}$ transforms according to a not necessarily irreducible
finite dimensional unitary representation $\mc{R}$; the corresponding index is denoted by $k$. 
The basis in this representation is chosen to be orthogonal and normalized to $1$.
The generators of the global YM Lie algebra in the representation $\mc{R}$ are denoted, 
after they are multiplied by $\ii$, by $(t_a)\indices{_k^l}$.
These matrices satisfy the commutation relations
$[t_a,t_b]=\ii f\indices{_{ab}^c} t_c$ of the Lie algebra of the global gauge group and are self-adjoint
(i.e.\ $t_a^{\dagger}=t_a$).

Under Lorentz transformations
$\psi\indices{_{\check{\nu}}^{\check{\rho}}_{\alpha k}}$ transforms according to a not necessarily irreducible finite dimensional real representation; the corresponding index is denoted by $\alpha$. A real representation is a direct sum of the irreducible
real representations $(\frac{l_1}{2},\frac{l_2}{2})\oplus (\frac{l_2}{2},\frac{l_1}{2})$, $l_1\ne l_2$, $l_1,l_2\in\NN$,
and $(\frac{l}{2},\frac{l}{2})$, $l\in\NN$, where $(\frac{l_1}{2},\frac{l_2}{2})$, $l_1,l_2\in\NN$, 
denote the finite dimensional irreducible representations of the Lorentz group. 
The simplest examples of these irreducible real representations are the Dirac spinor representation 
$(\frac{1}{2},0)\oplus (0,\frac{1}{2})$
and the Minkowski representation $(\frac{1}{2},\frac{1}{2})$.
A representation $(\frac{l_1}{2},\frac{l_2}{2})$ is fermionic if $l_1+l_2$ is odd and bosonic if 
$l_1+l_2$ is even. $(\frac{1}{2},0)$ and $(0,\frac{1}{2})$ are the Weyl spinor representations.
$\psi\indices{_{\check{\nu}}^{\check{\rho}}_{\alpha k}}$ can be taken to be number valued, 
but if it transforms according to a fermionic representation, then it is also reasonable to take it to be anticommuting.
The formulas written in this section are valid for both options.

The generalized Dirac conjugation is defined as
\beq
\label{eq.dc}
\bar{\psi}\indices{_{\check{\nu}}^{\check{\rho}\beta k}}= 
\psi\indices{_{\check{\nu}}^{\check{\rho}}_{\alpha k}^*}\epsilon^{\alpha\beta}\ ,
\eeq
where ${}^*$ denotes complex conjugation\footnote{For products of anticommuting quantities complex conjugation includes a reversal of the order of the factors.} and $\epsilon^{\alpha\beta}$ is a nondegenerate matrix 
with the properties 
$\epsilon^{\alpha\beta *}= \epsilon^{\beta\alpha}$ and 
$(L_{\bar{\mu}\bar{\nu}})\indices{_\delta^{\alpha *}}  \epsilon^{\delta\beta} +  (L_{\bar{\mu}\bar{\nu}})\indices{_\delta^{\beta}}  \epsilon^{\alpha\delta} = 0$. The latter property is needed for the Lorentz covariance of the Dirac conjugation, and the $L_{\bar{\mu}\bar{\nu}}$ appearing in it denote the generators of the Lorentz group in the representation according to which $\psi\indices{_{\check{\nu}}^{\check{\rho}}_{\alpha k}}$ transforms.
An $\epsilon^{\alpha\beta}$ matrix with these properties can be found for any finite dimensional real representation of the Lorentz group. If $\psi\indices{_{\check{\nu}}^{\check{\rho}}_{\alpha k}}$ transforms as a Dirac spinor and one uses e.g.\ the Weyl basis for Dirac spinors, then 
$\epsilon^{\alpha\beta}$ can be taken to be the Dirac gamma matrix $\gamma^{\bar{0}}$. 
For the purpose of the present paper it is not necessary to specify $\epsilon^{\alpha\beta}$ 
explicitly for the other representations. In general it would be necessary to insert into (\ref{eq.dc}) a matrix similar to $\epsilon^{\alpha\beta}$ for the YM indices as well, 
but with the special choice of basis in $\mc{R}$ mentioned above this matrix is the unit matrix. 

The restrictions on the possible indices of the matter field above are done in order to avoid writing large formulas.
Nevertheless, it is straightforward to extend the various formulas to the cases when $\psi$ transforms under Lorentz transformations
according to a nonreal representation (e.g.\ according to the Weyl spinor representations),
or when $\psi$ has several indices corresponding to Lorentz transformations. 

We also assume for simplicity that $\hat{L}$ is invariant under YM gauge transformations and local Lorentz transformations and 
transforms as a scalar function under diffeomorphisms. 
In this case the simplest and most natural choice for $K^\mu$ in the symmetry condition (\ref{eq.b4})
is $K^\mu = 0$ for YM gauge transformations and local Lorentz transformations 
and $K^\mu =- h^\mu L$ for diffeomorphisms. 

It should be noted that in the literature the matter fields are often regarded as sections of suitable vector bundles over $M$. 
In a coordinate based formalism it is usually assumed that a trivialization is fixed for these vector bundles over the coordinate patch $U$ that is under consideration. 

It is useful to introduce the following notation: 
\beq
\label{eq.not1}
P\indices{^{\mu\check{\nu}}_{\check{\rho}}^{\alpha k}} =  \frac{\partial \hat{L}}{\partial D_\mu \psi\indices{_{\check{\nu}}^{\check{\rho}}_{\alpha k}}} \qquad
\tilde{P}\indices{^{\mu\check{\nu}}_{\check{\rho}\alpha k}} =  \frac{\partial \hat{L}}{\partial D_\mu \bar{\psi}\indices{_{\check{\nu}}^{\check{\rho}\alpha k}}} 
\eeq
\beq
E\indices{^{\check{\nu}}_{\check{\rho}}^{\alpha k}} = \frac{1}{\sqrt{-g}}\frac{\delta L}{\delta \psi\indices{_{\check{\nu}}^{\check{\rho}}_{\alpha k}}}\qquad
\tilde{E}\indices{^{\check{\nu}}_{\check{\rho}\alpha k}} = \frac{1}{\sqrt{-g}}\frac{\delta L}{\delta \bar{\psi}\indices{_{\check{\nu}}^{\check{\rho}\alpha k}}}
\eeq
\bea
\mbb{A}^{\mu\lambda\nu}_i & = &
(P\indices{^{\mu\dots \lambda\dots}_{\check{\rho}}^{\alpha k}}
+P\indices{^{\lambda\dots\mu\dots}_{\check{\rho}}^{\alpha k}})\psi\indices{_{\dots}^\nu_{\dots}^{\check{\rho}}_{\alpha k}}
-
P\indices{^{\nu\dots\lambda\dots}_{\check{\rho}}^{\alpha k}}\psi\indices{_{\dots}^\mu_{\dots}^{\check{\rho}}_{\alpha k}}\nonumber\\
&& +(\tilde{P}\indices{^{\mu\dots \lambda\dots}_{\check{\rho}\alpha k}}
+\tilde{P}\indices{^{\lambda\dots\mu\dots}_{\check{\rho}\alpha k}})\bar{\psi}\indices{_{\dots}^\nu_{\dots}^{\check{\rho}\alpha k}}
-
\tilde{P}\indices{^{\nu\dots\lambda\dots}_{\check{\rho}\alpha k}}\bar{\psi}\indices{_{\dots}^\mu_{\dots}^{\check{\rho}\alpha k}}
\label{eq.not3}
\eea
\bea
\mbb{B}^{\mu\lambda\nu}_i & = & 
(P\indices{^{\mu\check{\nu}}_\dots^\lambda_\dots^{\alpha k}}
-P\indices{^{\lambda\check{\nu}}_\dots^\mu_\dots^{\alpha k}})\psi\indices{_{\check{\nu}}^{\dots\nu\dots}_{\alpha k}}
+
P\indices{^{\nu\check{\nu}}_\dots^\lambda_\dots^{\alpha k}}\psi\indices{_{\check{\nu}}^{\dots\mu\dots}_{\alpha k}}\nonumber\\
&& +(\tilde{P}\indices{^{\mu\check{\nu}}_\dots^\lambda_{\dots\alpha k}}
-\tilde{P}\indices{^{\lambda\check{\nu}}_\dots^\mu_{\dots\alpha k}})\bar{\psi}\indices{_{\check{\nu}}^{\dots\nu\dots\alpha k}}
+
\tilde{P}\indices{^{\nu\check{\nu}}_\dots^\lambda_{\dots\alpha k}}\bar{\psi}\indices{_{\check{\nu}}^{\dots\mu\dots\alpha k}}
\label{eq.not4}
\eea
\beq
\label{eq.not5}
\mbb{Q}^{\mu\bar{\delta}\bar{\lambda}}= \frac{1}{2} P\indices{^{\mu\check{\nu}}_{\check{\rho}}^{\alpha k}}
\psi\indices{_{\check{\nu}}^{\check{\rho}}_{\beta k}}(L^{\bar{\delta}\bar{\lambda}})\indices{_\alpha^\beta}
-\frac{1}{2} \tilde{P}\indices{^{\mu\check{\nu}}_{\check{\rho}\alpha k}}
\bar{\psi}\indices{_{\check{\nu}}^{\check{\rho}\beta k}}(L^{\bar{\delta}\bar{\lambda}})\indices{_\beta^\alpha}
\eeq
\makebox[\textwidth][s]{In (\ref{eq.not3}), for example, $P\indices{^{\mu\dots \lambda\dots}_{\check{\rho}}^{\alpha k}}$
at the beginning of the right hand side
means}\\
$P\indices{^{\mu\nu_1\nu_2\dots\nu_{i-1}\lambda \nu_{i+1}\dots \nu_n}_{\check{\rho}}^{\alpha k}}$, 
and similar notation is used subsequently. 

$\mbb{Q}^{\mu\bar{\delta}\bar{\lambda}}$ is antisymmetric in the last two indices.
The reality of $L$ implies that $P\indices{^{\mu\check{\nu}}_{\check{\rho}}^{\alpha k}}$ is the Dirac conjugate of $\tilde{P}\indices{^{\mu\check{\nu}}_{\check{\rho}\alpha k}}$
and
$E\indices{^{\check{\nu}}_{\check{\rho}}^{\alpha k}}$ is the Dirac conjugate of 
$\tilde{E}\indices{^{\check{\nu}}_{\check{\rho}\alpha k}}$
if 
$\psi\indices{_{\check{\nu}}^{\check{\rho}}_{\alpha k}}$ is an even field and 
$P\indices{^{\mu\check{\nu}}_{\check{\rho}}^{\alpha k}}$
is $-1$ times the Dirac conjugate of $\tilde{P}\indices{^{\mu\check{\nu}}_{\check{\rho}\alpha k}}$ 
and
$E\indices{^{\check{\nu}}_{\check{\rho}}^{\alpha k}}$ is $-1$ times the Dirac conjugate of $\tilde{E}\indices{^{\check{\nu}}_{\check{\rho}\alpha k}}$
if 
$\psi\indices{_{\check{\nu}}^{\check{\rho}}_{\alpha k}}$ is anticommuting.

\subsubsection{The Euler--Lagrange currents}

As a preliminary step one determines the currents $j_V^\mu$, $j_A^\mu$ and $j_\psi^\mu$ for arbitrary variations of the 
tetrad, the Yang--Mills vector potential and the matter field.
One finds
\beq
\label{eq.vm3}
j^\mu_A=0\ ,
\eeq
\beq
j^\mu_{\psi} = 
\sqrt{-g}\bigl( P\indices{^{\mu\check{\nu}}_{\check{\rho}}^{\alpha k}}\delta \psi\indices{_{\check{\nu}}^{\check{\rho}}_{\alpha k}}
+\tilde{P}\indices{^{\mu\check{\nu}}_{\check{\rho}\alpha k}}\delta \bar{\psi}\indices{_{\check{\nu}}^{\check{\rho}\alpha k}}\bigr)\ ,
\label{eq.vm5}
\eeq
and
\beq
\label{eq.vm11}
j_V^\mu = j_g^\mu + j_w^\mu 
\eeq
\beq
\label{eq.vm12}
j_g^\mu = \sqrt{-g}\biggl[-\frac{1}{2}\sum_{i=1}^n \mbb{A}_i^{\mu(\nu\lambda)}
+\frac{1}{2}\sum_{i=1}^m \mbb{B}_i^{\mu(\nu\lambda)}
+ \mbb{Q}^{(\nu\lambda)\mu}\biggr]\gamma_{\nu\lambda}
\eeq
\beq
\label{eq.vm13}
j_w^\mu = \sqrt{-g}\, \mbb{Q}^{\mu\nu\lambda}w_{[\nu\lambda]}
\eeq
$\gamma_{\mu\nu}=\delta g_{\mu\nu} = 2w_{(\mu\nu)}$, $w_{\mu\nu}=V_{\nu\bar{\mu}}\delta V_\mu^{\bar{\mu}}$.
The quantities $\mbb{A}_i^{\mu\nu\lambda}$, $\mbb{B}_i^{\mu\nu\lambda}$ and $\mbb{Q}_i^{\mu\nu\lambda}$ in 
(\ref{eq.vm12}) and (\ref{eq.vm13}) come from the variation of the connection.

With the above formulas at hand one can proceed to determining the Noether currents and the superpotentials for the various gauge symmetries. For this it is also necessary to know the variation of the matter field under the gauge transformations. 
The Noether currents $J^\mu$ are obtained as $J^\mu =  j_V^\mu +j_\psi^\mu -K^\mu$, 
specializing $\delta A_\mu^a$, $\delta V_\mu^{\bar{\mu}}$ and
$\delta \psi\indices{_{\check{\nu}}^{\check{\rho}}_{\alpha k}}$ in the above formulas for 
$j_V^\mu$ and $j_\psi^\mu$
according to the particular transformation rules (\ref{eq.ym1})-(\ref{eq.diff2}), (\ref{eq.ym3}), (\ref{eq.ll3}) and (\ref{eq.diff3}).
$J_\Phi^\mu \equiv J_\psi^\mu$ is $j_\psi^\mu -K^\mu$. 
Using (\ref{eq.ym3}), (\ref{eq.ll3}) or (\ref{eq.diff3}) one can calculate $\mc{B}_{\psi}^\mu$, then 
$I^\mu$ is obtained as $I^\mu = J^\mu + \mc{B}_A^\mu + \mc{B}_V^\mu + \mc{B}_{\psi}^\mu$, 
and finally $\Sigma^{\mu\nu}$ is given by (\ref{eq.i2}).
This general procedure for determining $I^\mu$ and $\Sigma^{\mu\nu}$ can be simplified
further in the present example in the following way: 
$I^\mu$ does not depend on higher than first derivatives of $\epsilon^\alpha$, i.e.\ 
$I^\mu=\epsilon^\alpha I_\alpha^\mu + \partial_\nu\epsilon^\alpha I_\alpha^{\mu\nu}$, 
thus according to (\ref{eq.i2}) $\Sigma^{\mu\nu} = \epsilon^\alpha I_\alpha^{[\mu\nu]}$.
This shows that $I_\alpha^{\mu\nu}$ completely determines $\Sigma^{\mu\nu}$ and $I^\mu$.
In particular, from $I^\mu = \partial_\nu\Sigma^{\mu\nu}$ we get $I_\alpha^\mu = \partial_\nu I_\alpha^{[\mu\nu]}$
and $I_\alpha^{\mu\nu} = I_\alpha^{[\mu\nu]}$. Furthermore, since the Bianchi currents and $K^\mu$ 
do not depend on the derivatives of $\epsilon^\alpha$, 
$I_\alpha^{\mu\nu} = J_\alpha^{\mu\nu} =  j_{V,\alpha}^{\mu\nu} +j_{\psi,\alpha}^{\mu\nu}$, i.e.\ 
\beq
\Sigma^{\mu\nu} = \epsilon^\alpha (j_{V,\alpha}^{\mu\nu} +j_{\psi,\alpha}^{\mu\nu}) \ .
\eeq
The notation $J_\alpha^{\mu\nu}$, $j_{V,\alpha}^{\mu\nu}$, $j_{\psi,\alpha}^{\mu\nu}$ is understood in the same way as 
$I_\alpha^{\mu\nu}$ and the index $\alpha$ is used here in the same sense as in Section \ref{sec.n}.

\subsubsection{Yang--Mills gauge symmetry}
\label{sec.ym}

The first order variation of the matter field under a YM gauge transformation
para-metrized by $\epsilon^a$ is
\beq
\label{eq.ym3}
\delta \psi\indices{_{\check{\nu}}^{\check{\rho}}_{\alpha k}}
=  \ii \kappa \epsilon^a  (t_a)\indices{_k^l} \psi\indices{_{\check{\nu}}^{\check{\rho}}_{\alpha l}} \ .
\eeq 
One finds that for YM gauge transformations the Noether current is
\beq
J^\mu =
\sqrt{-g}\, \epsilon^a \mc{J}^\mu_a\ ,
\eeq
which is obtained 
by specializing $\delta\psi_k$ in (\ref{eq.vm5}) according to (\ref{eq.ym3}).
Since $K^\mu= j_A^\mu = j_V^\mu =0$, 
$J^\mu = J_\psi^\mu = j_\psi^\mu$.
$J^\mu$ does not depend on the derivatives of $\epsilon^a$, therefore 
$I^\mu$ and $\Sigma^{\mu\nu}$ are $0$, and the Bianchi current $\mc{B}_{\chi}^\mu$ 
is identical to $-J^\mu$ even if the matter field does not satisfy its EL equation. 
The Bianchi currents $\mc{B}_{A}^\mu$ and $\mc{B}_{V}^\mu$ are given in (\ref{eq.ymb3a}), and 
$\mc{B}_{\psi}^\mu = 0$.

\subsubsection{Local Lorentz symmetry}
\label{sec.ll}

The first order variation of the matter field  
under a local Lorentz transformation is
\beq 
\label{eq.ll3}
\delta \psi\indices{_{\check{\nu}}^{\check{\rho}}_{\alpha k}}
=  \omega^{\bar{\mu}\bar{\nu}}(L_{\bar{\mu}\bar{\nu}})\indices{_\alpha^\beta}
\psi\indices{_{\check{\nu}}^{\check{\rho}}_{\beta k}}\ .
\eeq 
$\mc{B}_\psi^\mu$ is obviously zero and in Section \ref{sec.app1} we saw that also $\mc{B}_V^\mu=\mc{B}_A^\mu=0$.
After some calculation one finds that $J^\mu=0$ as well, thus $I^\mu = 0$.  
On the other hand, $J_\psi^\mu$ is not zero:
\beq
J_\psi^\mu = 2\sqrt{-g}\, \mbb{Q}^{\mu\bar{\nu}\bar{\eta}}\omega_{\bar{\nu}\bar{\eta}}\ . 
\eeq
A nonzero tetrad is not invariant under any Lorentz transformation, therefore $J_\psi^\mu$ is 
generally not conserved.
Nevertheless, $J_\psi^\mu$ has an important role when the tetrad and the vector potential are invariant under a 
combined transformation (see Section \ref{sec.mixedsymm}).

\subsubsection{Diffeomorphism symmetry}
\label{sec.diff}

The first order variation of the matter field under a diffeomorphism generated by $h^\mu$ is 
the Lie derivative
\bea
\delta \psi\indices{_{\check{\nu}}^{\check{\rho}}_{\alpha k}} &  = &  
-h^\mu \partial_\mu \psi\indices{_{\check{\nu}}^{\check{\rho}}_{\alpha k}}\nonumber \\
&& \hspace{-1.8cm} -\partial_{\nu_1} h^\mu \psi\indices{_{\mu\nu_2\dots\nu_n}^{\check{\rho}}_{\alpha k}}
-\partial_{\nu_2} h^\mu \psi\indices{_{\nu_1\mu\nu_3\dots\nu_n}^{\check{\rho}}_{\alpha k}}
-\dots
-\partial_{\nu_n} h^\mu \psi\indices{_{\nu_1\dots\nu_{n-1}\mu}^{\check{\rho}}_{\alpha k}}\nonumber\\
&& \hspace{-1.8cm} +\partial_\mu h^{\rho_1}\psi\indices{_{\check{\nu}}^{\mu\rho_2\dots\rho_m}_{\alpha k}}
+ \partial_\mu h^{\rho_2}\psi\indices{_{\check{\nu}}^{\rho_1\mu\rho_3\dots\rho_m}_{\alpha k}}
+\dots
+ \partial_\mu h^{\rho_m}\psi\indices{_{\check{\nu}}^{\rho_1\dots\rho_{m-1}\mu}_{\alpha k}}\ .
\label{eq.diff3}
\eea
It is straightforward to obtain explicit expressions for $J^\mu$ and $J^\mu_\psi$,  
but we do not write them out here, since they are not enlightening. 
The Bianchi currents are (\ref{eq.bva}) and
\bea 
\mc{B}_{\psi}^\mu & = & \sqrt{-g}\bigl[
- E\indices{^{\mu\nu_2\nu_3\dots}_{\check{\rho}}^{\alpha k}} h^\lambda \psi\indices{_{\lambda\nu_2\nu_3\dots}^{\check{\rho}}_{\alpha k}}
- E\indices{^{\nu_1\mu\nu_3\dots}_{\check{\rho}}^{\alpha k}} h^\lambda \psi\indices{_{\nu_1\lambda\nu_3\dots}^{\check{\rho}}_{\alpha k}}
-\dots\nonumber\\
&& +E\indices{^{\check{\nu}}_{\lambda\rho_2\rho_3\dots}^{\alpha k}}h^\lambda \psi\indices{_{\check{\nu}}^{\mu\rho_2\rho_3\dots}_{\alpha k}}
+E\indices{^{\check{\nu}}_{\rho_1\lambda\rho_3\dots}^{\alpha k}}h^\lambda
\psi\indices{_{\check{\nu}}^{\rho_1\mu\rho_3\dots}_{\alpha k}}+\dots\nonumber\\
&& +[\tilde{E}\bar{\psi}]\bigr]\ ,
\label{eq.diff4}
\eea
where $[\tilde{E}\bar{\psi}]$ represents the complex conjugate of the previous terms. \newpage
$I^\mu$ is found to be
\bea
I^\mu & = & \sqrt{-g}\Bigl[
\nabla_\nu [
2(\mbb{Q}^{(\mu\lambda)\nu}h_\lambda - \mbb{Q}^{(\nu\lambda)\mu}h_\lambda)
- \mbb{Q}^{\lambda\mu\nu}h_\lambda]\nonumber\\
&& -\frac{1}{2}\sum_{i=1}^n\nabla_\nu [P\indices{^{\mu\dots\nu\dots}_{\check{\rho}}^{\alpha k}}\psi\indices{_{\dots}^\lambda_{\dots}^{\check{\rho}}_{\alpha k}}h_\lambda
-P\indices{^{\nu\dots\mu\dots}_{\check{\rho}}^{\alpha k}}\psi\indices{_{\dots}^\lambda_{\dots}^{\check{\rho}}_{\alpha k}}h_\lambda]\nonumber\\
&& +\frac{1}{2}\sum_{i=1}^n\nabla_\nu[P\indices{^{\mu\dots\lambda\dots}_{\check{\rho}}^{\alpha k}}\psi\indices{_{\dots}^\nu_{\dots}^{\check{\rho}}_{\alpha k}}h_\lambda
-P\indices{^{\nu\dots\lambda\dots}_{\check{\rho}}^{\alpha k}}\psi\indices{_{\dots}^\mu_{\dots}^{\check{\rho}}_{\alpha k}}h_\lambda]\nonumber\\
&& +\frac{1}{2}\sum_{i=1}^n\nabla_\nu[P\indices{^{\lambda\dots\mu\dots}_{\check{\rho}}^{\alpha k}}\psi\indices{_{\dots}^\nu_{\dots}^{\check{\rho}}_{\alpha k}}h_\lambda
-P\indices{^{\lambda\dots\nu\dots}_{\check{\rho}}^{\alpha k}}\psi\indices{_{\dots}^\mu_{\dots}^{\check{\rho}}_{\alpha k}}h_\lambda]\nonumber\\
&& -\frac{1}{2}\sum_{i=1}^m \nabla_\rho [P\indices{^{\mu\check{\nu}}_\dots^\rho_\dots^{\alpha k}}\psi\indices{_{\check{\nu}}^{\dots\lambda\dots}_{\alpha k}}h_\lambda
-P\indices{^{\rho\check{\nu}}_\dots^\mu_\dots^{\alpha k}}\psi\indices{_{\check{\nu}}^{\dots\lambda\dots}_{\alpha k}}h_\lambda
] \nonumber \\
&& +\frac{1}{2}\sum_{i=1}^m \nabla_\rho [P\indices{^{\mu\check{\nu}}_\dots^\lambda_\dots^{\alpha k}}\psi\indices{_{\check{\nu}}^{\dots\rho\dots}_{\alpha k}}h_\lambda
-P\indices{^{\rho\check{\nu}}_\dots^\lambda_\dots^{\alpha k}}\psi\indices{_{\check{\nu}}^{\dots\mu\dots}_{\alpha k}}h_\lambda
] \nonumber\\
&& +\frac{1}{2}\sum_{i=1}^m \nabla_\rho [P\indices{^{\lambda\check{\nu}}_\dots^\mu_\dots^{\alpha k}}\psi\indices{_{\check{\nu}}^{\dots\rho\dots}_{\alpha k}}h_\lambda
-P\indices{^{\lambda\check{\nu}}_\dots^\rho_\dots^{\alpha k}}\psi\indices{_{\check{\nu}}^{\dots\mu\dots}_{\alpha k}}h_\lambda
] \nonumber\\
&& + [\tilde{P}\bar{\psi}] \Bigr]\ ,
\label{eq.Im}
\eea
where $[\tilde{P}\bar{\psi}]$ represents the complex conjugate of the $P\psi$ type terms in the previous six lines. 
$\Sigma^{\mu\nu}$ is obtained by omitting the $\nabla$ operators in (\ref{eq.Im}) and changing the $\rho$ index to $\nu$ where necessary. 

It is interesting to note that if the matter field is scalar with respect to diffeomorphisms and local Lorentz transformations,
then $\mc{B}_{\psi}^\mu = 0$ and $I^\mu=0$, therefore the Bianchi current $\mc{B}_{\chi}^\mu$ 
coincides with $-J^\mu$ even if 
the matter field does not satisfy its EL equations.
$\mc{B}_{\psi}^\mu=0$ holds even if the matter field is scalar only with respect to diffeomorphisms.

\subsubsection{Mixed symmetries}
\label{sec.mixedsymm}

A general infinitesimal gauge transformation is a combination of the special transformations discussed in the previous subsections
and is characterized by a triple $(h^\mu, \omega^{\bar{\mu}\bar{\nu}}, \epsilon^a)$.
For such transformations $\delta V_\mu^{\bar{\mu}}$, $\delta A_\mu^a$ and 
$\delta \psi\indices{_{\check{\nu}}^{\check{\rho}}_{\alpha k}}$,
the Bianchi currents, the Noether currents, 
$I^\mu$ and $\Sigma^{\mu\nu}$  
are just the sums of the relevant expressions 
for the special transformations. 

A common case when combined transformations are important is when a spacetime has complete rotation symmetry ($SO(3)$).
Since it is not possible to choose a tetrad that is invariant under 
the entire rotation group (understood as a subgroup of the group of diffeomorphisms), 
some of the rotation symmetry transformations of the tetrad necessarily involve local Lorentz transformations,
and this has to be taken into account when one calculates Noether currents.

\subsubsection{The Dirac field}
\label{sec.dirac}

An important example of a matter Lagrangian is the Dirac Lagrangian
\beq
L=\sqrt{-g}\biggl[\frac{1}{2}g^{\mu\nu}(\bar{\psi}\ii\gamma_\mu D_\nu\psi
-D_\nu\bar{\psi}\ii\gamma_\mu\psi)-m\bar{\psi}\psi\biggr]\ ,
\eeq
where $\psi$ denotes the Dirac field. As usual, the Dirac spinor indices are omitted, 
and the indices related to YM gauge group representations are also suppressed. 
$\gamma^\mu$
are defined as
$\gamma^\mu=V^\mu_{\bar{\mu}}\gamma^{\bar{\mu}}$,
where $\gamma^{\bar{\mu}}$ are the standard Dirac gamma matrices in Minkowski space (see e.g.\ \cite{PS}).
The generators of the Lorentz group $L^{\bar{\nu}\bar{\lambda}}$ in the Dirac representation are
$L^{\bar{\mu}\bar{\nu}} = \frac{1}{2}\sigma^{\bar{\mu}\bar{\nu}}$, where
$\sigma^{\bar{\mu}\bar{\nu}} =
\frac{1}{2}[\gamma^{\bar{\mu}},\gamma^{\bar{\nu}}]$.

$I^\mu$ (see (\ref{eq.Im})) takes a relatively simple form for the Dirac field. 
Since the Dirac field does not have vector and covector indices, only the $\mbb{Q}$ terms are present in (\ref{eq.Im}).
The quantities $\mbb{A}^{\mu\lambda\nu}$ and 
$\mbb{B}^{\mu\lambda\nu}$ introduced in (\ref{eq.not3}) and (\ref{eq.not4}) are also absent. 
The quantity $\mbb{Q}^{\mu\bar{\delta}\bar{\lambda}}$ introduced in (\ref{eq.not5}) is
$\mbb{Q}^{\mu\bar{\delta}\bar{\lambda}} = \frac{1}{8}(\bar{\psi}\ii\gamma^\mu \sigma^{\bar{\delta}\bar{\lambda}}\psi +
\bar{\psi}\sigma^{\bar{\delta}\bar{\lambda}}\ii\gamma^\mu \psi)$.
$\gamma^\mu \sigma^{\delta\lambda} + \sigma^{\delta\lambda} \gamma^\mu = \gamma^\lambda\gamma^\mu\gamma^\delta - 
\gamma^\delta\gamma^\mu\gamma^\lambda$, therefore
$\mbb{Q}^{\mu\delta\lambda}$ can also be written as 
\beq
\mbb{Q}^{\mu\delta\lambda} = \frac{1}{8}\ii\bar{\psi} (\gamma^\lambda\gamma^\mu\gamma^\delta - 
\gamma^\delta\gamma^\mu\gamma^\lambda)\psi\ .
\eeq
It is not difficult to see that 
$\mbb{Q}^{(\nu\lambda)\mu} = 0$, thus 
\beq
\label{eq.Idirac}
I^\mu = -\sqrt{-g}\, \nabla_\nu (\mbb{Q}^{\lambda\mu\nu}h_\lambda)
\eeq
in the case of the Dirac field.
This result for $I^\mu$ was also found in \cite{Fletcher} 
in the special case of the electromagnetic field as YM gauge field. 
In \cite{Fletcher} the approach to treating spinor fields was different from the one applied in this paper.

For a diffeomorphism symmetry under which the tetrad and the vector potential are invariant
the canonical Noether current $J_\psi^\mu$ is found to be
$-\sqrt{-g}\frac{1}{2}(\bar{\psi}\ii\gamma^\mu \nabla_\nu \psi - \nabla_\nu\bar{\psi}\ii\gamma^\mu \psi)h^\nu - h^\mu L$.
For the Einstein-Hilbert energy-momentum tensor one finds the expression
$T^{\mu\nu}= \frac{1}{4}(
\bar{\psi}\ii\gamma^\mu D^\nu\psi + \bar{\psi}\ii\gamma^\nu D^\mu\psi
-D^\mu\bar{\psi}\ii\gamma^\nu\psi -D^\nu\bar{\psi}\ii\gamma^\mu\psi)$, and
$\mc{J}^\mu_a = -\kappa\bar{\psi}\gamma^\mu t_a \psi$. 

In \cite{TG2} it was found in an ad hoc manner that subtracting $\sqrt{-g}\,\nabla_\nu (\mbb{Q}^{\lambda\mu\nu}h_\lambda)$
from the canonical Noether current 
$-J_\psi^\mu$
gives (\ref{eq.0}), if the Dirac equation is satisfied.
Here we have been able to derive this within a general formalism. 
(Note that in \cite{TG2} the YM gauge field was the electromagnetic field and 
a factor $-1$ was included in the definition of the Noether currents.)

\section{Matter fields in the presence of fixed gravitational and scalar fields}
\label{sec.5}

In this example we discuss only the partial Bianchi identity and the Bianchi current.
The matter Lagrangian is again assumed to have the standard form $L=\sqrt{-g}\hat{L}$, and
$\hat{L}$ is assumed to be a local function of the tetrad, the real scalar field $\phi$, and the other fields. 
$\hat{L}$ is also assumed to be invariant under local Lorentz transformations and to transform as a scalar function under diffeomorphisms. The total gauge symmetry group in this example is thus the group generated by diffeomorphisms 
and local Lorentz transformations. 
The nature of the matter fields does not need to be specified.

The first order variation of $\phi$ under a diffeomorphism is
$\delta \phi = -h^\mu \partial_\mu \phi$. Using this property
one obtains $B_{\phi\mu}=-\sqrt{-g}\mc{J}_\phi\partial_\mu\phi$ for 
for the Bianchi expression $B_{\phi\mu}$, with  
$\mc{J}_\phi = \frac{1}{\sqrt{-g}}\frac{\delta L}{\delta \phi}$. $B_{V\mu}$ is given by (\ref{eq.bvmu}).
Taking into consideration $\mathbb{T}_{\bar{\mu}\bar{\nu}} = 0$ (see (\ref{eq.lb1})), 
the partial Bianchi identity $B_{V\mu}+B_{\phi\mu}=0$ is thus 
\beq
\label{eq.4x}
B_\mu = \sqrt{-g}(-\nabla_\nu T\indices{^\nu_\mu}-\mc{J}_\phi\partial_\mu\phi )=0\ .
\eeq
The Bianchi current $\mc{B}_\phi^\mu$ is obviously zero, therefore the Bianchi current $\mc{B}_\chi^\mu$ is  
$\sqrt{-g}\,T\indices{^\mu_\nu}h^\nu$, i.e.\ it takes the same form as in the absence of the fixed scalar field.
$\mc{B}_\chi^\mu$ is conserved if $h^\mu$ is a Killing field, the first order variation of $\phi$
with respect to the diffeomorphisms generated by $h^\mu$ is also zero, and the matter fields satisfy their EL equations.
The conservation of  $\mc{B}_\chi^\mu$ can be derived from (\ref{eq.4x}) and from the symmetry properties of the metric and the scalar field:
$\nabla_\mu (T\indices{^\mu_\nu}h^\nu) = \nabla_\mu T\indices{^\mu_\nu}h^\nu +  T\indices{^\mu_\nu}  \nabla_\mu h^\nu$, and 
here on the right hand side the second term is zero in virtue of the Killing equation, whereas the first term can be rewritten as 
$\nabla_\mu T\indices{^\mu_\nu}h^\nu = - \mc{J}_\phi h^\nu \partial_\nu \phi$ using (\ref{eq.4x}). 
$\mc{J}_\phi h^\nu \partial_\nu \phi$ is indeed zero if $\delta \phi=0$.

It is interesting to note that if only the scalar field is fixed, then the partial Bianchi identity becomes 
$B_\mu = \sqrt{-g}(-\mc{J}_\phi\partial_\mu\phi )=0$ (and now $\mc{J}_\phi$ is obtained from the total Lagrangian), 
as is also found in \cite{BS}. This identity has the remarkable consequence that either $\phi$ is constant 
or it also satisfies its EL equation.

\section{Conclusion}
\label{sec.concl}

In this paper we extended the standard construction of conserved currents 
associated with spacetime symmetries for matter fields propagating 
in fixed curved spacetime to general gauge theories, 
without any restriction on the order of the derivatives of the fields that may appear in the Lagrangian.
In particular we showed that if in a Lagrangian field theory that has gauge symmetry
in the general Noetherian sense
some of the elementary fields are fixed 
and are invariant under an infinitesimal
gauge transformation, then there exists a current, which we called Bianchi current, that is 
analogous to the current $\sqrt{-g}\,T\indices{^\mu_\nu}h^\nu$ used in general relativity and is conserved if the non-fixed fields
satisfy their Euler--Lagrange equations. The conservation of this current can be seen as a consequence 
of the symmetry of the fixed fields and of an identity, which we called partial Bianchi identity,
that is analogous to $\nabla_\mu T^{\mu\nu}=0$ and follows from the 
gauge symmetry of the Lagrangian. 
We also showed that the Noether current associated with the symmetry of the fixed fields obtained by applying Noether's first theorem differs from the Bianchi current by the sum of an identically conserved current and a term that vanishes if the non-fixed fields satisfy their Euler--Lagrange equations.
We gave explicit formulas for the Bianchi current and for the other quantities appearing in these results, 
so they can be calculated in any particular model.
If the total Lagrangian can be split in the same way as in general relativity to a `matter' and a `gravitational' part, 
so that the latter depends only on the fixed fields, then the construction can be applied to the `matter' part separately,
as is done in general relativity in the standard case.

As example we discussed first the case of general matter fields propagating in backgrounds consisting of a 
gravitational and a Yang--Mills field. We found (\ref{eq.db2}) 
as the generalization of the current $\sqrt{-g}\,T\indices{^\mu_\nu}h^\nu$.
The extension of (\ref{eq.db2}) to the case when the Yang--Mills part of the background, i.e.\ the vector potential, is invariant under the diffeomorphisms generated by $h^\mu$ only up to (Yang--Mills) gauge transformations is (\ref{eq.dyb}).
As the generalization of the property $\nabla_\mu T^{\mu\nu}=0$ we found the Lorentz law (\ref{eq.db1}).
This means that the Lorentz law is found to hold  
in arbitrary gravitational and Yang--Mills background,
if the matter fields satisfy their Euler--Lagrange equations and the matter Lagrangian 
has diffeomorphism, local Lorentz and Yang--Mills gauge symmetry.
For local Lorentz transformations and Yang--Mills gauge transformations we found the partial Bianchi identities (\ref{eq.lb1}) and (\ref{eq.ymb1}). The Bianchi current for a Yang--Mills gauge symmetry of the fixed fields is (\ref{eq.ymb3b}).  
Under not very restrictive assumptions on the type of the matter fields and on the form of the matter Lagrangian 
we investigated the difference between the Bianchi currents and the Noether currents. 
In the case of Yang--Mills gauge symmetries these currents coincide even if the dynamical fields do not satisfy 
their Euler--Lagrange equations. 
In the case of diffeomorphism symmetries the Bianchi currents are generally not identical with the Noether currents, 
except if the matter fields are scalar fields, as the known results for zero fixed Yang--Mills field also indicate.
We obtained the formulas (\ref{eq.diff4}) and (\ref{eq.Im}) for the characterization of the difference between the Bianchi and Noether currents. In the case of the Dirac field  (\ref{eq.Im}) reduces to (\ref{eq.Idirac}).

If the requirement of the Yang--Mills symmetry of the Lagrangian is omitted but 
fixed covector fields are nevertheless present, then the generalization of $\nabla_\mu T^{\mu\nu}=0$ becomes
(\ref{eq.db3}) instead of (\ref{eq.db1}), 
whereas the Bianchi current for a diffeomorphism symmetry has the unchanged form (\ref{eq.db2}).
 
The second example was the case of fields propagating in backgrounds consisting of a gravitational and a real scalar field. 
As the generalization of $\nabla_\mu T^{\mu\nu}=0$ we found (\ref{eq.4x}), whereas the Bianchi current turned out to have the same form,
$\sqrt{-g}\,T\indices{^\mu_\nu}h^\nu$, as in the absence of the fixed scalar field.

The construction presented in this paper can be applied in a very wide variety of models, for example in metric-affine gravitation theory or in other extended models of gravitation coupled with Yang--Mills type gauge fields and matter fields. It would be interesting to see if the Bianchi currents that can be constructed in these models can be used to obtain results similar to those in \cite{NQV}. 
$p$-form field theory and other higher spin gauge theories are further examples that could be investigated.

Although in the examples that we discussed one of the fixed fields was the gravitational field, 
one can also apply the construction in cases when the gravitational field is not among the fixed fields. 
Examples of partial Bianchi identities for such cases have already been given in \cite{BS}.

\section*{Acknowledgments}

The author is supported by an MTA Lend\"ulet grant and by the NKFIH grant K116505.

\appendix

\renewcommand{\theequation}{\Alph{section}.\arabic{equation}} 
\setcounter{equation}{0}

\section{Auxiliary formulas and remarks}
\label{sec.A}

In Section \ref{sec.eymm} the complete gauge group is the group generated by the diffeomorphisms, 
the local Lorentz transformations and the YM gauge transformations. 
The covariant derivative of the 
matter field $\psi\indices{_{\check{\nu}}^{\check{\rho}}_{\alpha k}}$---introduced at the beginning of Section \ref{sec.app2}---for this gauge group is given by the formula
\beq
\label{eq.covd}
D_\mu \psi\indices{_{\check{\nu}}^{\check{\rho}}_{\alpha k}}  = 
\nabla_\mu  \psi\indices{_{\check{\nu}}^{\check{\rho}}_{\alpha k}}
+ S\indices{_\alpha^\beta_\mu} \psi\indices{_{\check{\nu}}^{\check{\rho}}_{\beta k}}
-\ii\kappa A_\mu^a (t_a)\indices{_k^l}  \psi\indices{_{\check{\nu}}^{\check{\rho}}_{\alpha l}}\ ,
\eeq
where $\nabla_\mu$ denotes the Levi--Civita covariant differentiation
corresponding to $g_{\mu\nu}$ and
$S\indices{_\alpha^\beta_\mu} = \frac{1}{2}(L^{\bar{\nu}\bar{\lambda}})\indices{_\alpha^\beta}\mc{S}_{\bar{\nu}\bar{\lambda}\mu}$,
$\mc{S}\indices{^{\bar{\lambda}}_{\bar{\eta}\mu}}=-V_{\bar{\eta}}^\nu\nabla_\mu V_\nu^{\bar{\lambda}}$.
$(L^{\bar{\nu}\bar{\lambda}})\indices{_\alpha^\beta}$ and $(t_a)\indices{_k^l}$
are the generators of the Lorentz group and of the global YM gauge group, respectively, 
in the representations according to which the matter field transforms. 
The second term on the right hand side of (\ref{eq.covd}) describes the action of $D_\mu$ on the Lorentz group related indices, 
and the third term gives the action of $D_\mu$ on the YM indices. 
More details concerning the second term can be found e.g.\ in \cite{W}.
One also applies $D_\mu$ as an operator to other fields that have the same types of indices as the matter field, 
even if they transform somewhat differently, as the vector potential $A_\mu^a$, for example.  

The local Lorentz transformations and the YM gauge transformations form two normal subgroups in the complete gauge group,
but the group of diffeomorphisms is not a normal subgroup either with respect to YM gauge transformations
or with respect to local Lorentz transformations. 
In accordance with this situation the Lie derivatives of those fields that are not scalar with respect to 
local Lorentz transformations or YM gauge transformations generally transform in a noncovariant manner
under these transformations. 

Since the subgroup of diffeomorphisms is not a normal subgroup in the complete gauge group, 
from purely group theoretical point of view the complete gauge group does not have a unique diffeomorphism
subgroup, rather it has many diffeomorphism subgroups conjugate to one another. 
Nevertheless, in a coordinate based formalism a definite diffeomorphism subgroup
becomes distinguished implicitly, which can be called the subgroup of diffeomorphisms.

\end{document}